\documentstyle[12pt]{article}

\begin{document}
\thispagestyle{empty}

\newcommand{\p}[1]{(\ref{#1})}
\newcommand{\be}{\begin{equation}}
\newcommand{\ee}{\end{equation}}
\newcommand{\sect}[1]{\setcounter{equation}{0}\section{#1}}

\newcommand{\vs}[1]{\rule[- #1 mm]{0mm}{#1 mm}}
\newcommand{\hs}[1]{\hspace{#1mm}}
\newcommand{\mb}[1]{\hs{5}\mbox{#1}\hs{5}}
\newcommand{\Db}{{\overline D}}
\newcommand{\bea}{\begin{eqnarray}}
\newcommand{\eea}{\end{eqnarray}}
\newcommand{\wt}[1]{\widetilde{#1}}
\newcommand{\und}[1]{\underline{#1}}
\newcommand{\ov}[1]{\overline{#1}}
\newcommand{\sm}[2]{\frac{\mbox{\footnotesize #1}\vs{-2}}
           {\vs{-2}\mbox{\footnotesize #2}}}
\newcommand{\prt}{\partial}
\newcommand{\eps}{\epsilon}

\newcommand{\R}{\mbox{\rule{0.2mm}{2.8mm}\hspace{-1.5mm} R}}
\newcommand{\Z}{Z\hspace{-2mm}Z}

\newcommand{\cd}{{\cal D}}
\newcommand{\cg}{{\cal G}}
\newcommand{\ck}{{\cal K}}
\newcommand{\cw}{{\cal W}}

\newcommand{\vj}{\vec{J}}
\newcommand{\vl}{\vec{\lambda}}
\newcommand{\vz}{\vec{\sigma}}
\newcommand{\vt}{\vec{\tau}}
\newcommand{\vw}{\vec{W}}
\newcommand{\poiss}{\stackrel{\otimes}{,}}

\def\l#1#2{\raisebox{.2ex}{$\displaystyle
  \mathop{#1}^{{\scriptstyle #2}\rightarrow}$}}
\def\r#1#2{\raisebox{.2ex}{$\displaystyle
 \mathop{#1}^{\leftarrow {\scriptstyle #2}}$}}

% REVUES POUR BIBLIO

\newcommand{\NP}[1]{Nucl.\ Phys.\ {\bf #1}}
\newcommand{\PL}[1]{Phys.\ Lett.\ {\bf #1}}
\newcommand{\NC}[1]{Nuovo Cimento {\bf #1}}
\newcommand{\CMP}[1]{Comm.\ Math.\ Phys.\ {\bf #1}}
\newcommand{\PR}[1]{Phys.\ Rev.\ {\bf #1}}
\newcommand{\PRL}[1]{Phys.\ Rev.\ Lett.\ {\bf #1}}
\newcommand{\MPL}[1]{Mod.\ Phys.\ Lett.\ {\bf #1}}
\newcommand{\BLMS}[1]{Bull.\ London Math.\ Soc.\ {\bf #1}}
\newcommand{\IJMP}[1]{Int.\ Jour.\ of\ Mod.\ Phys.\ {\bf #1}}
\newcommand{\JMP}[1]{Jour.\ of\ Math.\ Phys.\ {\bf #1}}
\newcommand{\LMP}[1]{Lett.\ in\ Math.\ Phys.\ {\bf #1}}

%\begin{document}
\renewcommand{\thefootnote}{\fnsymbol{footnote}}
\newpage
\setcounter{page}{0}
\pagestyle{empty}
\begin{flushright}
{October 1997}\\
{SISSA 131/97/EP}\\
{hep-th/9710118}
\end{flushright}
\vs{8}
\begin{center}
{\LARGE {\bf The $N=2$ supersymmetric Toda lattice}}\\[0.6cm]
{\LARGE {\bf hierarchy  and matrix models}}\\[1cm]

\vs{8}

{\large L. Bonora$^{a,1}$ and A. Sorin$^{b,2}$}
{}~\\
\quad \\
{\em ~$~^{(a)}$ International School for Advanced Studies
(SISSA/ISAS),}\\
{\em Via Beirut 2, 34014 Trieste, Italy}\\
{\em {~$~^{(b)}$ Bogoliubov Laboratory of Theoretical Physics,
JINR,}}\\
{\em 141980 Dubna, Moscow Region, Russia}~\quad\\

\end{center}
\vs{8}

\centerline{ {\bf Abstract}}
\vs{4}

We propose a new integrable $N=2$ supersymmetric Toda lattice hierarchy 
which may be relevant for constructing a supersymmetric one--matrix model.
We define its first two Hamiltonian structures, the
recursion operator and Lax--pair representation. We provide partial evidence 
for the existence of an infinite-dimensional $N=2$ superalgebra 
of its flows. We study its 
bosonic limit and introduce new Lax--pair representations
for the bosonic Toda lattice hierarchy. Finally we discuss the relevance
this approach for constructing
$N=2$ supersymmetric generalized Toda lattice hierarchies.

\vfill
{\em E-Mail:\\
1) bonora@sissa.it\\
2) sorin@thsun1.jinr.dubna.su }
\newpage

\pagestyle{plain}
\renewcommand{\thefootnote}{\arabic{footnote}}
\setcounter{footnote}{0}

\section{Introduction.}
In this paper we construct the $N=2$ version of the 
Toda lattice hierarchy. We write down the Hamiltonians, the recursion operator 
and the bosonic as well as the fermionic flows.

There are several reasons to study this problem. First, one would like to extend 
also to discrete integrable hierarchies the program of $N=2$ 
supersymmetrization which has proven very successful for differential
hierarchies. Secondly: it is well--known that from a bosonic integrable 
hierarchy of the Toda type one can construct a differential hierarchy; it is
interesting to see whether this hold true also for the $N=2$ supersymmetric
extension. Finally, there have been a few attempts to find a supersymmetric 
generalization of the one--matrix model. It is arguable whether these attempts
have been successful. We believe it is worth trying a different course,
based on the remark that most information concerning exactly solvable
matrix models is contained in the underlying integrable hierarchies. The idea is
to find first the $N=2$ supersymmetric extension of the integrable
hierarchy that characterizes the one--matrix model, and then try to reconstruct 
the features of the supersymmetric model which is at the origin of it. Here
we have completed the first step in this direction.

Behind these motivations is the basically unanswered question concerning the 
relevance of $N=2$ supersymmetric extensions of integrable hierarchies 
particularly in connection with 2D cohomological field theories. It is
well--known that such theories can be obtained as twisted
2D $N=2$  supersymmetric theories. On the other hand many integrable hierarchies
correspond to cohomological field theories (for example models corresponding
to the A and D series). The same hierarchies are also likely to admit an 
$N=2$ extension. Although we do not know enough yet about supersymmetric
hierarchies, it is likely that the above is not an accidental coincidence.
This paper adds some further evidence in this direction.

The original motivation for this paper was to find an $N=2$ supersymmetric
extension of the Toda hierarchy underlying hermitean one--matrix model. This
is defined by the semi--infinite matrix 
\begin{eqnarray}
Q= \sum_j (E_{jj+1} + a_j E_{jj}+ b_j E_{jj-1})\nonumber
\end{eqnarray}
and by the flows
\begin{eqnarray}
\frac {\partial Q}{\partial t_k} = [Q_+^k , Q],\nonumber
\end{eqnarray}
where the subscript + means the upper triangular part of a matrix including
the main diagonal, and $E_{ij}$ is the matrix with entries $(E_{ij})_{kl}=
\delta_{ik}\delta_{jl}$. Actually our final result is more general than this.
It also includes, for example, the $N=2$ versions of finite lattice 
hierarchies. 

It remains for us to point out that
our construction of the $N=2$ Toda lattice hierarchy presented here
hinges upon some recent results on $N=2$ hierarchies,
\cite{dls,s,ls}, 
and to describe the content of this paper. In section 2 we introduce the
first two Hamiltonian structures and the lowest lying (bosonic and
fermionic) Hamiltonians. In section 3 we compute the recursion
operator and, starting
from it, in section 4 we write down recursion formulas for all the flows
as well as explicit formulas for the first two sets of flows. Section 5 
shows that our hierarchy actually possesses $N=2$ supersymmetry. We conjecture
that it is characterized by a more general non--abelian algebra.
Then we the possible reductions of our general
hierarchy: one leads to the lattice hierarchy characteristic of one--matrix
model; the others are reductions over finite lattices. In section 6 we
show how we arrived at the formulation of the present hierarchy starting
from the $N=2$ supersymmetric Nonlinear Schr\"odinger (NLS)
hierarchy and the f--Toda equations. We also give
indications of how it is possible to extend the construction of this paper 
to generalized $N=2$ Toda lattice hierarchies. In section 7 we
discuss Lax--pair 
formulations of our hierarchy and in section 8 their bosonic limits. An 
appendix is devoted to the Lagrangian formulation of the first
flow.

\section{ Hamiltonian structure of the $N=2$ super
Toda lattice hierarchy.}
The $N=2$ supersymmetric Toda lattice hierarchy comprises four
different classes of bi-Hamiltonian flows. The first is a
hierarchy of commuting bosonic flows with the evolution times $t_l$ 
($l \in {\bf N}$) and Hamiltonians $H_l$ which are in involution,
and in the bosonic limit they reproduce the flows and Hamiltonians of the
corresponding bosonic Toda lattice hierarchy. In
addition to $H_l$, there exist one more bosonic and two fermionic series
of Hamiltonians $U_l$, $S_l$ and ${\overline S}_l$, respectively, which
are integrals of the $t_l$-flows generated by Hamiltonians $H_l$ and
commute with them. In other words the flows generated by them represent 
symmetries of the $t_l$-flows. In general, all Hamiltonians as well as 
their flows form a nonabelian algebra. Altogether, these data may be
considered as a  general definition of the extended integrable
hierarchy we are looking for.

In this section we describe the bi-Hamiltonian structure of the hierarchy;
later on we will present the flows,
their properties, algebra and different representations.

Let us introduce first some notations:
\begin{eqnarray}
{\cal H}_{a,l} \equiv \{U_l, S_l, {\overline S}_l,H_l \}, \quad {\tau}_{a,l}
\equiv \{q_l, {\theta}_l, {\overline {\theta}}_l , t_l \},
\quad a,b=1,..., 4,
\label{def0}
\end{eqnarray}
\begin{eqnarray}
O_{A,i} \equiv \{
b_{i}, a_{i}, {\beta}_{i}, {\overline {\beta}}_{i}, {\alpha}_{i},
{\overline {\alpha}}_{i} \}, \quad A,B=1, ..., 6, \quad 
i \in {\bf Z},
\label{def}
\end{eqnarray}
where ${\tau}_{a,l}$ are evolution times corresponding to
Hamiltonians ${\cal H}_{a,l}$. More precisely, ${\theta}_l$ and ${\overline
{\theta}}_l$ ($t_l$ and $q_l$) are Grassmann odd (even) variables; $b_{i}$
and $a_{i}$ (${\beta}_i, {\overline {\beta}}_{i}, {\alpha}_i$ and
${\overline {\alpha}}_{i}$) are bosonic (fermionic) fields which depend on
all evolution times.

Lattice Hamiltonians \p{def0} are represented in the following
general form:
\begin{eqnarray}
{\cal H}_{a,l}=\sum_{j=-\infty}^{\infty} h_{a,l,j}.
\label{hamlat}
\end{eqnarray}
As a consequence of our definition of the hierarchy, any Hamiltonian
density $h_{a,l,j}$ is expected to satisfy an equation with respect to the
evolution time $t_l$, which has the form of a lattice conservation law,
\begin{eqnarray}
{\textstyle{\partial\over\partial t_l}}h_{a,l,j}= f_j-f_{j-1}
\equiv (\Delta f)_j, \quad \sum_{j=-\infty}^{\infty}(\Delta f)_j=0 
\Rightarrow 
{\textstyle{\partial\over\partial t_l}}{\cal H}_{a,l}=0, 
\label{dens}
\end{eqnarray}
where $f_j$ is a polynomial of the lattice fields $O_{A,i}$ \p{def}.
In what follows we call the operator $\Delta$ lattice
derivative and assume a suitable boundary conditions for the fields
$O_{A,i}$ in order for the last equality in eqs. \p{dens} to be satisfied.

The first nontrivial fermionic, $S_l$ and ${\overline S}_l$,
and bosonic, $U_l$ and $H_l$, Hamiltonians are:
\begin{eqnarray}
&& U_0=\sum_{j=-\infty}^{\infty} {\ln b_{j}}; \quad
U_1=\sum_{j=-\infty}^{\infty} (-\frac{{\beta}_j
{\overline {\beta}}_j}{b_{j}} +
{\alpha}_j \sum_{i=-\infty}^{j-1} {\overline {\alpha}}_i), \nonumber\\
&& S_1 = \sum_{j=-\infty}^{\infty} {\alpha}_j, \quad
S_2 = \sum_{j=-\infty}^{\infty} ({\beta}_j + {\alpha}_j
\sum^{j-1}_{i=-\infty}(a_{i} + \frac{{\beta}_i
{\overline {\beta}}_i}{b_{i}})), \nonumber\\
&& {\overline S}_1=-\sum_{j=-\infty}^{\infty} {\overline {\alpha}}_j,\quad
{\overline S}_2 = \sum_{j=-\infty}^{\infty}
({\overline {\beta}}_j - {\overline {\alpha}}_j\sum_{i=j+1}^{\infty}(a_{i-1}+
\frac{{\beta}_i{\overline {\beta}}_i}{b_{i}})), \nonumber\\
&& H_1 = \sum_{j=-\infty}^{\infty} ( a_{j} +
\frac{{\beta}_j{\overline {\beta}}_j}{b_{j}}), \quad
H_2 = \sum_{j=-\infty}^{\infty} (\frac{1}{2} a_{j}^{2} + b_{j} -
{\beta}_j{\overline {\alpha}}_j + {\alpha}_j{\overline {\beta}}_j).
\label{hams}
\end{eqnarray}
Their dimensions in length can be chosen as $[S_{l}] = [{\overline S}_{l}]=
{-l+1/2}$ and $[H_{l}] = [U_l]=-l$. Then the dimensions of
all the fields are completely fixed and are defined by:
$[b_i]=-2$, $[a_i]=-1$, $[{\beta}_i]=
[{\overline {\beta}}_i]=-3/2$ and $[{\alpha}_i]=
[{\overline {\alpha}}_i]=-1/2$. Let us remark the local character
of the densities corresponding to the Hamiltonians $H_l$ \p{hams}.

A bi--Hamiltonian system of equations can be represented in the following
general form:
\begin{eqnarray}
&& {\textstyle{\partial\over\partial {\tau}_{a,l}}} O_{A,i}=
\{ {\cal H}_{a,l+1}, O_{A,i} \}_1 =\{ {\cal H}_{a,l}, O_{A,i} \}_2  
\nonumber\\
&& \equiv \sum_{j=-\infty}^{\infty}\sum_{B=1}^{6}(J_{1})_{AB,ij}
{\textstyle{\delta \over\delta O_{B,j}}} {\cal H}_{a,l+1}=
\sum_{j=-\infty}^{\infty}\sum_{B=1}^{6} (J_{2})_{AB,ij}
{\textstyle{\delta \over\delta O_{B,j}}} {\cal H}_{a,l},
\label{biham}
\end{eqnarray}
where $J_1$ and $J_2$ are supermatrices of the first and second
Hamiltonian structures
\begin{eqnarray}
(J_p)_{AB,ij} = - (-1)^{d_{A}d_{B}} (J_p)_{BA,ji}
\equiv -(-1)^{d_Ad_{{\cal H}_a}}\{ O_{A,i}, O_{B,j} \}_p, \quad p=1,2,
\label{hamstrp}
\end{eqnarray}
and the brackets $\{,\}_p$ are corresponding graded Poisson brackets
with the properties
\begin{eqnarray}
&& \{ O_{A,i}, O_{B,j} \}_p = - (-1)^{d_{A}d_{B}}
\{ O_{B,j}, O_{A,i} \}_p, \nonumber\\
&& \{ O_{A,i}, O_{B,j} O_{C,k}\}_p =
\{ O_{A,i}, O_{B,j}\}_p O_{C,k} + (-1)^{d_{A}d_{B}}
O_{B,j}\{ O_{A,i},O_{C,k}\}_p,
\label{prop}
\end{eqnarray}
and satisfying the graded Jacobi identities
\begin{eqnarray}
&& (-1)^{d_{A}d_{C}}\{ \{O_{A,i}, O_{B,j}\}_p, O_{C,k}\}_p +
(-1)^{d_{B}d_{A}}\{ \{O_{B,i}, O_{C,j}\}_p, O_{A,k}\}_p \nonumber\\
&& +(-1)^{d_{C}d_{B}}\{ \{O_{C,i}, O_{A,j}\}_p, O_{B,k}\}_p = 0.
\label{Jacobi}
\end{eqnarray}
Here, the $d_A$ is the Grassmann parity of the fields $O_{A,i}$, $d_{A}=0$
($1$) for bosonic (fermionic) fields, and the $d_{{\cal H}_a}$
is the Grassmann parity of the Hamiltonians ${\cal H}_{a,l}$ \p{def0}.

The explicit form for the $J_1$ and $J_2$ can be obtained using their
definitions \p{hamstrp} and the following explicit expressions for the
first,
\begin{eqnarray}
&& \{ b_i, a_j \}_1=b_i (-\delta_{i,j} +\delta_{i,j+1}), \nonumber\\
&& \{ a_i,{\beta}_j \}_1= {\beta}_j
\delta_{i,j}, \nonumber\\ && \{ a_i,{\overline {\beta}}_j \}_1=-
{\overline
{\beta}}_j \delta_{i,j-1}, \nonumber\\ && \{ {\beta}_i, {\overline
{\beta}}_j \}_1=-b_{j} \delta_{i,j}, \nonumber\\ && \{
{\alpha}_i,{\overline {\alpha}}_j \}_1 = -\delta_{i,j} +\delta_{i,j+1},
\label{hamstr1}
\end{eqnarray}
and for the second,
\begin{eqnarray}
&& \{ b_i, b_j \}_2=b_i b_j (\delta_{i,j+1} - \delta_{i,j-1}), \nonumber\\
&& \{ b_i, a_j \}_2=b_i a_j (-\delta_{i,j}+\delta_{i,j+1}), \nonumber\\
&& \{ b_i,{\beta}_j \}_2 = b_i{\beta}_j \delta_{i,j+1}, \nonumber\\
&& \{ b_i,{\overline {\beta}}_j \}_2=-b_i {\overline {\beta}}_j
\delta_{i,j-1}, \nonumber\\
&& \{ b_i,{\alpha}_j \}_2 = b_i{\alpha}_j \delta_{i,j}, \nonumber\\
&& \{ b_i,{\overline {\alpha}}_j \}_2=-b_i{\overline {\alpha}}_j
\delta_{i,j}, \nonumber\\
&& \{ a_i, a_j \}_2=b_i \delta_{i,j+1}-b_j \delta_{i,j-1}, \nonumber\\
&& \{ a_i,{\beta}_j \}_2 =
a_i {\beta}_j \delta_{i,j} - b_j{\alpha}_j \delta_{i,j-1}, \nonumber\\
&& \{ a_i,{\overline {\beta}}_j \}_2=-a_i {\overline {\beta}}_j
\delta_{i,j-1} - b_{i} {\overline {\alpha}}_j \delta_{i,j}, \nonumber\\
&& \{ a_i,{\alpha}_j \}_2 = {\beta}_j \delta_{i,j},  \nonumber\\
&& \{ a_i,{\overline {\alpha}}_j \}_2 = {\overline {\beta}}_j
\delta_{i,j-1}, \nonumber\\
&& \{ {\beta}_i, {\overline {\beta}}_j \}_2=
-{\beta}_i {\overline {\beta}}_j  \delta_{i,j-1}, \nonumber\\
&& \{ {\beta}_i, {\overline {\alpha}}_j \}_2=
-{\beta}_i{\overline {\alpha}}_j\delta_{i,j}+b_{i} \delta_{i,j+1},\nonumber\\
&&\{{\overline {\beta}}_i, {\alpha}_j \}_2= -{\alpha}_j {\overline {\beta}}_i
\delta_{i,j}-b_{i}\delta_{i,j-1}, \nonumber\\
&& \{ {\alpha}_i , {\overline {\alpha}}_j \}_2 =
\frac{{\beta}_i{\overline {\beta}}_i}{b_{i}} \delta_{i,j} +
a_{j} \delta_{i,j+1},
\label{hamstr2}
\end{eqnarray}
Poisson brackets structures, where only nonzero brackets are written down.

The algebra of the Poisson brackets \p{hamstr1} possesses the
discrete inner automorphism $\sigma_j$ defined by 
\begin{eqnarray}
&& \sigma_j b_i {\sigma}^{-1}_j= b_{j-i}, \quad
\sigma_j a_i {\sigma}^{-1}_j=-a_{j-i-1} , \nonumber\\
&& \sigma_j {\beta}_i {\sigma}^{-1}_j={\overline {\beta}}_{j-i}, \quad
\sigma_j {\overline {\beta}}_i {\sigma}^{-1}_j= {\beta}_{j-i}, \nonumber\\
&& \sigma_j {\alpha}_i {\sigma}^{-1}_j={\overline {\alpha}}_{j-i}, \quad
\sigma_j {\overline {\alpha}}_i {\sigma}^{-1}_j= {\alpha}_{j-i}.
\label{auto1}
\end{eqnarray}
Under the action of these transformations the overall signs of all
Poisson brackets of the algebra \p{hamstr2} are reversed. Due to this
as well as to the property $\sigma_j^2=1$, one can conclude that the
${\sigma}_j$-transformations are involution transformations of the
algebra \p{hamstr2}. 

Now taking into account the following transformation
properties of the Hamiltonians \p{hams}
\begin{eqnarray}
\sigma_j\{U_l, S_l, {\overline S}_l,H_l \}{\sigma}^{-1}_j=
(-1)^{l}\{U_l, {\overline S}_l, S_l, H_l \}
\label{auto2}
\end{eqnarray}
and defining the transformation properties of their evolution parameters
\p{def0} by the formula
\begin{eqnarray}
\sigma_j \{q_l, {\theta}_l, {\overline {\theta}}_l ,t_l\}{\sigma}^{-1}_j=
(-1)^{l-1}\{q_l, {\overline {\theta}}_l, {\theta}_l, t_l \},
\label{auto3}
\end{eqnarray}
one easily recognizes that the bi-Hamiltonian system \p{biham}
also possesses the involution ${\sigma}_j$ \p{auto1}, \p{auto3}.

A remark is in order: the appearance of $b_j$ in the denominator 
both in the Poisson
brackets \p{hamstr2} and Hamiltonians \p{hams}, is an artifact of the basis.  
One can  avoid it if instead of the fields ${\beta}_i,{\overline
{\beta}}_i$ one uses the following ones: ${\beta}_i\mapsto
b_i{\beta}_i$, ${\overline {\beta}}_i\mapsto {\overline {\beta}}_i$ (or
${\beta}_i\mapsto {\beta}_i$, ${\overline {\beta}}_i\mapsto b_i{\overline
{\beta}}_i$). However, we prefer to deal with the old fields
${\beta}_i,{\overline {\beta}}_i$  as the transformation
properties \p{auto1} with respect to the involution ${\sigma}_j$ are simpler.

\section{Recursion operator of the $N=2$ super
Toda lattice hierarchy.}
One can check by direct, but rather tedious calculations that
the graded Jacobi identities \p{Jacobi} are satisfied for the algebras
\p{hamstr1}--\p{hamstr2}. Moreover, they are also satisfied for their sum
${\mu}_1\{,\}_1+{\mu}_1\{,\}_2$, where ${\mu}_1$ and ${\mu}_2$ are
arbitrary parameters. Thus, the Hamiltonian structures $J_1$ and $J_2$ are
compatible, i.e. they form a Hamiltonian pair which can be used to
construct the hereditary recursion operator\footnote{For details
concerning integrable Hamiltonian systems, see, e.g., the review \cite{af} and
references therein.} $R_{AB,ij}$ according to the following general rule:
\begin{eqnarray}
&&R_{AB,ij} \equiv
\sum_{k=-\infty}^{\infty}\sum_{C=1}^{6} (J_2)_{AC,ik} (J_1^{-1})_{CB,kj},
\label{recop}
\end{eqnarray}
where $(J_1^{-1})_{CB,kj}$ is the inverse matrix of the first
Hamiltonian structure,
\begin{eqnarray}
&&\sum_{k=-\infty}^{\infty}\sum_{C=1}^{6} (J_1)_{AC,ik} (J_1^{-1})_{CB,kj}
=\sum_{k=-\infty}^{\infty}\sum_{C=1}^{6} (J_1^{-1})_{AC,ik} (J_1)_{CB,kj}
= \delta_{A,B} \delta_{i,j}, \nonumber\\
&& (J_1^{-1})_{11,ij} =\frac{1}{b_{i}b_{j}}
( \frac{{\beta}_{i} {\overline {\beta}}_{i}}{b_{i}}
\sum_{k \geq 1 } \delta_{i+k,j} -
\frac{{\beta}_{j} {\overline {\beta}}_{j}}{b_{j}}
\sum_{k \geq 1 } \delta_{i-k,j}), \nonumber\\
&& (J_1^{-1})_{12,ij} = - (J_1^{-1})_{21,ji} =
\frac{1}{b_{i}} \sum_{k \geq 1 } \delta_{i-k,j}, \nonumber\\
&& (-1)^{d_{{\cal H}_a}}(J_1^{-1})_{13,ij} = (J_1^{-1})_{31,ji} =
-\frac{{\overline {\beta}}_{j}}{b_{i}b_{j}} \sum_{k \geq 1 }
\delta_{i-k+1,j}, \nonumber\\
&&(-1)^{d_{{\cal H}_a}} (J_1^{-1})_{14,ij} =
(J_1^{-1})_{41,ji}=\frac{{\beta}_{j}}{b_{i}b_{j}}
\sum_{k \geq 1 } \delta_{i-k,j}, \nonumber\\
&& (J_1^{-1})_{34,ij} =  (J_1^{-1})_{43,ji}
= (-1)^{d_{{\cal H}_a}}\frac{1}{b_{i}} \delta_{i,j}, \nonumber\\
&& (J_1^{-1})_{56,ji} =(J_1^{-1})_{65,ij} =
(-1)^{d_{{\cal H}_a}}\sum_{k \geq 1 } \delta_{i-k+1,j}.
\label{hamstr-1}
\end{eqnarray}
Here, only nonzero matrix elements are presented.

Another important consequence of \p{biham} and
the compatibility of the Hamiltonian structures are the following
involution properties
\begin{eqnarray}
\{{\cal H}_{a,l}, {\cal H}_{a,m}\}_2=
\{{\cal H}_{a,l}, {\cal H}_{a,m}\}_1=0
\label{hamsinvol}
\end{eqnarray}
of the Hamiltonians \p{def0}.

The recursion operator \p{recop} can be represented by $(6\infty) \times
(6\infty)$ supermatrix with the following general structure:
\begin{eqnarray}
R=\pmatrix{B_{(2\infty) \times (2\infty)}, &
(-1)^{d_{{\cal H}_a}}F_{(2\infty) \times 4(\infty)}\cr
(-1)^{d_{{\cal H}_a}}F_{(4\infty) \times (2\infty)}, &
B_{(4\infty) \times (4\infty)} \cr},
\label{supmatr}
\end{eqnarray}
where $B_{(n\infty) \times (n\infty)}$ ($F_{(n\infty) \times (m\infty)}$)
is a boson- (fermion-) valued square (rectangular) matrix of dimension
$(n\infty) \times (n\infty)$ ($(n\infty) \times (m\infty)$) which is the
same both for the case of bosonic and fermionic Hamiltonians \p{def0},
\p{hams}. The dependence of the recursion operator on the
Grassmann nature of the Hamiltonians appears in \p{supmatr} only
via the factors $(-1)^{d_{{\cal H}_a}}$. Substituting eqs. \p{hamstr2} and
\p{hamstr-1} into eq. \p{recop}, one can obtain the following explicit
expressions for its entries:
\begin{eqnarray}
&& B_{11,ij}=(a_{i-1} -\frac{{\beta}_{i} {\overline {\beta}}_{i}}{b_{i}})
\delta_{i,j} + \frac{b_{i}}{b_{j}} ( a_{i-1} - a_{i}) \sum_{k \geq 1 }
\delta_{i,j-k}, \quad
B_{12,ij} = b_{i} (\delta_{i,j} + \delta_{i,j+1}), \label{recopb1}\\
&& B_{21,ij}=\frac{1}{b_{j}} (b_{i} \sum_{k \geq 1 } \delta_{i,j-k+1}
-({\alpha}_{i+1} {\overline {\beta}}_{i+1}
+ {\beta}_{i} {\overline {\alpha}}_{i}) \sum_{k \geq 1 } \delta_{i,j-k}-
b_{i+1} \sum_{k \geq 1 } \delta_{i,j-k-1}), \quad
B_{22,ij}=a_i \delta_{i,j},\nonumber 
\end{eqnarray}
\begin{eqnarray}
&& F_{13,ij}=-{\overline {\beta}}_{j} \delta_{i,j}, \quad
F_{14,ij} = {\beta}_{j} \delta_{i,j}, \quad
F_{15,ij}=b_{i} {\overline {\alpha}}_{i}
\sum_{k \geq 1 } \delta_{i,j+k-1}, \quad
F_{16,ij} = -b_{i} {\alpha}_{i}\sum_{k \geq 1 }\delta_{i,j-k+1},
\nonumber\\
&&F_{23,ij}={\overline {\alpha}}_{i} \delta_{i,j}, \quad
F_{24,ij}={\alpha}_{j} \delta_{i,j-1}, \quad
F_{25,ij}=-{\overline {\beta}}_{i+1}\sum_{k \geq 1 } \delta_{i,j+k-2}, \quad
F_{26,ij} = -{\beta}_{i} \sum_{k \geq 1 } \delta_{i,j-k+1},\nonumber  
\end{eqnarray}
and
\begin{eqnarray}
&&F_{31,ij}=\frac{1}{b_{j}} ( b_{i} {\alpha}_{i}\sum_{k \geq 1 }
\delta_{i,j-k+1} -a_i{\beta}_{i} \sum_{k \geq 1 } \delta_{i,j-k} -
{\beta}_{i} \frac{{\beta}_{j} {\overline {\beta}}_{j}}{b_{j}}
\sum_{k \geq 1 } \delta_{i,j+k-1}), \quad
\nonumber\\
&& F_{41,ij}=\frac{1}{b_{j}} ( b_{i} {\overline {\alpha}}_{i}
\sum_{k \geq 1 } \delta_{i,j-k} + a_{i-1}{\overline {\beta}}_{i}
\sum_{k \geq 1 } \delta_{i,j-k+1} +{\overline {\beta}}_{i}
\frac{{\beta}_{j}{\overline {\beta}}_{j}}{b_{j}}
\sum_{k \geq 1 } \delta_{i,j+k}), \nonumber\\
&&F_{32,ij}={\beta}_{i} \sum_{k \geq 1} \delta_{i,j+k-1}, \quad\quad
F_{42,ij}=-{\overline {\beta}}_{i}\sum_{k \geq 1}\delta_{i,j+k+1},\nonumber\\
&& F_{51,ij}=\frac{1}{b_{j}}({\beta}_{i-1} \sum_{k \geq 1 }
\delta_{i,j-k+1} -{\beta}_{i} \sum_{k \geq 1 }\delta_{i,j-k} -
{\alpha}_{i} \frac{{\beta}_{j} {\overline {\beta}}_{j}}{b_{j}}
\sum_{k \geq 1 } \delta_{i,j+k}), \quad
F_{52,ij}={\alpha}_{i}\sum_{k \geq 1 } \delta_{i,j+k}, \nonumber\\
&& F_{61,ij}=\frac{1}{b_{j}}({\overline {\beta}}_{i+1}\sum_{k \geq 1 }
\delta_{i,j-k}-{\overline {\beta}}_{i}\sum_{k \geq 1 } \delta_{i,j-k+1}+
{\overline {\alpha}}_{i}\frac{{\beta}_{j} {\overline {\beta}}_{j}}{b_{j}}
\sum_{k \geq 1 } \delta_{i,j+k-1}), \quad
F_{62,ij}=-{\overline {\alpha}}_{i}\sum_{k \geq 1}
\delta_{i,j+k},
\nonumber
\end{eqnarray}
\begin{eqnarray}
&& B_{33,ij}= - \frac{{\beta}_{i} {\overline {\beta}}_{j}}{b_{j}}
\sum_{k \geq 1} \delta_{i,j+k-1}, \quad
B_{34,ij}= \frac{{\beta}_{i} {\beta}_{j}}{b_{j}}
\sum_{k \geq 1} \delta_{i,j+k-1},  \nonumber\\
&&B_{35,ij}={\beta}_{i} {\overline {\alpha}}_i\sum_{k \geq 1}
\delta_{i,j+k-1}-b_{i} \sum_{k \geq 1} \delta_{i,j+k}, \quad  
B_{43,ij}=\frac{{\overline {\beta}}_{i}{\overline {\beta}}_{j}}{b_{j}}
\sum_{k \geq 1}\delta_{i,j+k}, \nonumber\\
&&B_{44,ij}=\frac{{\beta}_{j} {\overline {\beta}}_{i}}{b_{j}}
\sum_{k \geq 1} \delta_{i,j+k}, \quad
B_{46,ij}={\alpha}_{i} {\overline {\beta}}_i\sum_{k \geq 1}
\delta_{i,j-k+1}+b_{i}\sum_{k \geq 1} \delta_{i,j-k}, \label{recopb2}\\
&&B_{53,ij}=\delta_{i,j+1} -\frac{{\alpha}_{i}
{\overline {\beta}}_{j}}{b_{j}}\sum_{k \geq 1 } \delta_{i,j+k}, \quad
B_{54,ij}=\frac{{\alpha}_{i} {\beta}_{j}}{b_{j}}
\sum_{k \geq 1 } \delta_{i,j+k}, \nonumber\\
&& B_{55,ij}=-\frac{{\beta}_{i} {\overline {\beta}}_{i}}{b_{i}}
\sum_{k \geq 1 } \delta_{i,j+k-1}-
a_{i-1} \sum_{k \geq 1 }\delta_{i,j+k},
\nonumber\\ &&B_{63,ij}= \frac{{\overline {\alpha}}_{i}{\overline
{\beta}}_{j}}{b_{j}} \sum_{k \geq 1 } \delta_{i,j+k-1}, \quad
B_{64,ij}= - \delta_{i,j-1} +\frac{{\beta}_{j} {\overline
{\alpha}}_{i}} {b_{j}}\sum_{k \geq 1 } \delta_{i,j+k-1}, \nonumber\\
&&B_{66,ij}=-\frac{{\beta}_{i} {\overline {\beta}}_{i}}
{b_{i}}\sum_{k \geq 1 } \delta_{i,j-k+1}- a_{i} \sum_{k \geq 1 }
\delta_{i,j-k}, \quad B_{36,ij}=R_{45,ij}=B_{56,ij}=B_{65,ij}=0. 
\nonumber
\end{eqnarray}

\section{Flows of the $N=2$ super Toda lattice hierarchy.}
The hierarchy starts with the flows corresponding to the evolution times
$q_0$, ${\theta}_1$, ${\overline {\theta}}_1$ and $t_1$, because
the Hamiltonians $U_0$, $S_1$, ${\overline S}_1$ and $H_1$ \p{hams}
lie in the center of the first Hamiltonian structure, i.e.,
\begin{eqnarray}
\{ U_0, O_{A,i} \}_1=\{ S_1, O_{A,i} \}_1 =
\{ {\overline S}_1, O_{A,i} \}_1=\{ H_1, O_{A,i} \}_1=0.
\label{start}
\end{eqnarray}
Taking into account eqs. \p{biham}, the relations \p{start} obviously
demonstrate that there are no flows with the evolution times $q_{-1}$,
${\theta}_0$, ${\overline {\theta}}_0$ and $t_0$. The first nontrivial
fermionic and bosonic flows with the times $q_0$, ${\theta}_1$,
${\overline {\theta}}_1$ and $t_1$ can be derived using the Hamiltonians
$U_0$, $S_1$, ${\overline S}_1$, $H_1$ (or $U_1$, $S_2$, ${\overline S}_2$,
$H_2$) \p{hams}, respectively, and the second (or first) Hamiltonian
structure \p{hamstr2} (\p{hamstr1}),
\begin{eqnarray}
&&{\textstyle{\partial\over\partial q_0}}b_j =0, \quad
{\textstyle{\partial\over\partial q_0}}a_j =0,\nonumber\\
&&{\textstyle{\partial\over\partial q_0}}{\beta}_j ={\beta}_j, \quad
{\textstyle{\partial\over\partial q_0}}{\overline {\beta}}_j
=-{\overline {\beta}}_j, \nonumber\\
&&{\textstyle{\partial\over\partial q_0}}{\alpha}_j={\alpha}_j,\quad
{\textstyle{\partial\over\partial q_0}}
{\overline {\alpha}}_j= -{\overline {\alpha}}_j,
\label{flow0}
\end{eqnarray}
\begin{eqnarray}
&&{\textstyle{\partial\over\partial {\theta}_1}}b_j=-b_j{\alpha}_{j},\quad
{\textstyle{\partial\over\partial {\theta}_1}}a_j=-{\beta}_j,\nonumber\\
&&{\textstyle{\partial\over\partial {\theta}_1}}{\beta}_j =0, \quad
{\textstyle{\partial\over\partial {\theta}_1}}{\overline {\beta}}_j
=-b_{j}-{\alpha}_j{\overline {\beta}}_j, \nonumber\\
&&{\textstyle{\partial\over\partial {\theta}_1}}{\alpha}_j=0,\quad
{\textstyle{\partial\over\partial {\theta}_1}}
{\overline {\alpha}}_j= a_{j} +
\frac{{\beta}_j{\overline {\beta}}_j}{b_{j}},
\label{flow11/2}
\end{eqnarray}
\begin{eqnarray}
&&{\textstyle{\partial\over\partial {\overline {\theta}}_1}}b_j =
-b_j{\overline {\alpha}}_{j}, \quad
{\textstyle{\partial\over\partial {\overline {\theta}}_1}}a_{j-1}=
{\overline {\beta}}_{j}, \nonumber\\
&&{\textstyle{\partial\over\partial {\overline {\theta}}_1}}{\beta}_j
=-b_{j}+{\beta}_j {\overline {\alpha}}_j,\quad
{\textstyle{\partial\over\partial {\overline {\theta}}_1}}
{\overline {\beta}}_j=0, \nonumber\\
&&{\textstyle{\partial\over\partial {\overline {\theta}}_1}}{\alpha}_j
=-a_{j-1} - \frac{{\beta}_j{\overline {\beta}}_j}{b_{j}}, \quad
{\textstyle{\partial\over\partial {\overline {\theta}}_1}}
{\overline {\alpha}}_j=0,
\label{flow21/2}
\end{eqnarray}
\begin{eqnarray}
&&{\textstyle{\partial\over\partial t_1}}b_j=b_j(a_{j}-a_{j-1}), \quad
{\textstyle{\partial\over\partial t_1}}a_j=b_{j+1} - b_{j}
+{\beta}_j{\overline {\alpha}}_j + {\alpha}_{j+1}
{\overline {\beta}}_{j+1}, \nonumber\\
&&{\textstyle{\partial\over\partial t_1}}{\beta}_j
=a_{j}{\beta}_j - b_{j}{\alpha}_{j}, \quad
{\textstyle{\partial\over\partial t_1}}{\overline {\beta}}_j
=-a_{j-1}{\overline {\beta}}_j - b_{j}{\overline {\alpha}}_j, \nonumber\\
&&{\textstyle{\partial\over\partial t_1}}{\alpha}_j
={\beta}_j - {\beta}_{j-1}, \quad
{\textstyle{\partial\over\partial t_1}}{\overline {\alpha}}_j
={\overline {\beta}}_j - {\overline {\beta}}_{j+1}.
\label{flow1}
\end{eqnarray}

The higher flows can be generated from flows \p{flow0}--\p{flow1}
using the recurrence relations
\begin{eqnarray}
{\textstyle{\partial\over\partial {\tau}_{a,l+1}}} O_{A,i} =
\sum_{j=-\infty}^{\infty}\sum_{B=1}^{6} R_{AB,ij}
{\textstyle{\partial\over\partial {\tau}_{a,l}}} O_{B,j} \equiv
(K_{a,l+1})_{A,i}.
\label{flows}
\end{eqnarray}
Substituting eqs. \p{def} and \p{supmatr}--\p{recopb2} into \p{flows},
one can obtain the following explicit expressions for them:
\begin{eqnarray}
{\textstyle{\partial\over\partial {\tau}_{a,l+1}}}b_i &=&
b_i[{\textstyle{\partial\over\partial {\tau}_{a,l}}}(a_i+a_{i-1}+
\frac{{\beta}_j{\overline {\beta}}_j}{b_{j}})+a_{i-1}
{\textstyle{\partial\over\partial {\tau}_{a,l}}}\sum_{j=i}^{\infty}
{\ln b_j}-a_{i} {\textstyle{\partial\over\partial {\tau}_{a,l}}}
\sum_{j=i+1}^{\infty}{\ln b_j} \nonumber\\
&+&(-1)^{d_{{\tau}_{a}}}{\overline {\alpha}}_i
{\textstyle{\partial\over\partial {\tau}_{a,l}}}
\sum_{j=-\infty}^{i} {\alpha}_j-(-1)^{d_{{\tau}_{a}}}{\alpha}_i
{\textstyle{\partial\over\partial {\tau}_{a,l}}}
\sum_{j=i}^{\infty}{\overline {\alpha}}_j], \nonumber\\
{\textstyle{\partial\over\partial {\tau}_{a,l+1}}}a_i &=&
{\textstyle{\partial\over\partial {\tau}_{a,l}}}
(b_{i+1}+b_{i}+ \frac{1}{2}a_{i}^{2}-
{\beta}_j{\overline {\alpha}}_j + {\alpha}_{j+1}{\overline {\beta}}_{j+1})
\nonumber\\
&-& (b_{i+1}-b_{i}+ {\beta}_j{\overline {\alpha}}_j + {\alpha}_{j+1}
{\overline {\beta}}_{j+1}){\textstyle{\partial\over\partial {\tau}_{a,l}}}
\sum_{j=i+1}^{\infty}{\ln b_j}\nonumber\\
&-&(-1)^{d_{{\tau}_{a}}}{\overline {\beta}}_{i+1}
{\textstyle{\partial\over\partial {\tau}_{a,l}}}
\sum_{j=-\infty}^{i} {\alpha}_j-(-1)^{d_{{\tau}_{a}}}{\beta}_i
{\textstyle{\partial\over\partial {\tau}_{a,l}}}
\sum_{j=i+1}^{\infty}{\overline {\alpha}}_j, \nonumber\\
{\textstyle{\partial\over\partial {\tau}_{a,l+1}}}{\beta}_i &=&
{\textstyle{\partial\over\partial {\tau}_{a,l}}}(b_i {\alpha}_{i})+
(-1)^{d_{{\tau}_{a}}}{\beta}_{i}{\textstyle{\partial\over\partial
{\tau}_{a,l}}} \sum_{j=-\infty}^{i}( a_{j}+
\frac{{\beta}_j{\overline {\beta}}_j}{b_{j}})\nonumber\\
&+&(-1)^{d_{{\tau}_{a}}}
(b_i {\alpha}_{i}-a_i{\beta}_i){\textstyle{\partial\over\partial
{\tau}_{a,l}}}\sum_{j=i+1}^{\infty}{\ln b_j}  
-(b_{i}-{\beta}_{i}
{\overline {\alpha}}_i) {\textstyle{\partial\over\partial {\tau}_{a,l}}}
\sum_{j=-\infty}^{i} {\alpha}_j, \nonumber\\
{\textstyle{\partial\over\partial {\tau}_{a,l+1}}}{\overline {\beta}}_i &=&
-{\textstyle{\partial\over\partial {\tau}_{a,l}}}(b_i{\overline \alpha}_{i})
-(-1)^{d_{{\tau}_{a}}}{\overline \beta}_{i}{\textstyle{\partial\over\partial
{\tau}_{a,l}}} \sum_{j=-\infty}^{i-1}( a_{j-1} +
\frac{{\beta}_j{\overline {\beta}}_j}{b_{j}})\nonumber\\
&+&(-1)^{d_{{\tau}_{a}}}
(b_i {\overline \alpha}_{i}+a_{i-1}
{\overline \beta}_i){\textstyle{\partial\over\partial {\tau}_{a,l}}}
\sum_{j=i}^{\infty}{\ln b_j} +(b_{i}+{\alpha}_{i}
{\overline {\beta}}_i) {\textstyle{\partial\over\partial {\tau}_{a,l}}}
\sum_{j=i}^{\infty} {\overline \alpha}_j, \nonumber\\
{\textstyle{\partial\over\partial {\tau}_{a,l+1}}}{\alpha}_i &=&
{\textstyle{\partial\over\partial {\tau}_{a,l}}}
({\beta}_{i-1}+a_{i-1} {\alpha}_{i})+
(-1)^{d_{{\tau}_{a}}}{\alpha}_{i}{\textstyle{\partial\over\partial
{\tau}_{a,l}}} \sum_{j=-\infty}^{i-1}( a_{j-1} +
\frac{{\beta}_j{\overline {\beta}}_j}{b_{j}})\nonumber\\
&-&( a_{i-1} +\frac{{\beta}_i{\overline {\beta}}_i}{b_{i}})
{\textstyle{\partial\over\partial {\tau}_{a,l}}}
\sum_{j=-\infty}^{i} {\alpha}_j  
 - (-1)^{d_{{\tau}_{a}}}{\beta}_i{\textstyle{\partial\over\partial
{\tau}_{a,l}}}\sum_{j=i+1}^{\infty}{\ln b_j}
+(-1)^{d_{{\tau}_{a}}}{\beta}_{i-1}{\textstyle{\partial\over\partial
{\tau}_{a,l}}}\sum_{j=i}^{\infty}{\ln b_j}, \nonumber\\
{\textstyle{\partial\over\partial {\tau}_{a,l+1}}}{\overline \alpha}_i &=&
-{\textstyle{\partial\over\partial {\tau}_{a,l}}}
(\overline {\beta}_{i+1}-a_{i} {\overline \alpha}_{i})-
(-1)^{d_{{\tau}_{a}}}{\overline \alpha}_{i}
{\textstyle{\partial\over\partial {\tau}_{a,l}}}
\sum_{j=-\infty}^{i}( a_{j} + \frac{{\beta}_j{\overline {\beta}}_j}{b_{j}})
- ( a_{i} +\frac{{\beta}_i{\overline {\beta}}_i}{b_{i}})
{\textstyle{\partial\over\partial {\tau}_{a,l}}}
\sum_{j=i}^{\infty} {\overline \alpha}_j \nonumber\\
&+&(-1)^{d_{{\tau}_{a}}}{\overline \beta}_{i+1}
{\textstyle{\partial\over\partial {\tau}_{a,l}}}\sum_{j=i+1}^{\infty}
{\ln b_j}-(-1)^{d_{{\tau}_{a}}}{\overline \beta}_{i}
{\textstyle{\partial\over\partial {\tau}_{a,l}}}
\sum_{j=i}^{\infty}{\ln b_j},
\label{recrel}
\end{eqnarray}
where $d_{{\tau}_{a}}$ is the Grassmann parity of the evolution time
${\tau}_{a,l}$. Everywhere in these expressions only the lattice
densities of the Hamiltonians $U_0$, $S_1$, ${\overline S}_1$ and $H_1$
\p{hams} appear under the sign of summation over the lattice points and
under the action of time derivatives. Remembering that the
$t_l$-derivative of any Hamiltonian density \p{dens} can be represented
as the lattice derivative of some local functions of the fields $O_{A,i}$
\p{def}, one can conclude that all $t_l$-flows and densities of their
Hamiltonians $H_l$ are local. This is, in general, not the case for
other
flows and Hamiltonians of the hierarchy under consideration.

Let us comment now about the effect of the involution $\sigma_j$ on the
flows.
One can check by direct calculations that, if the flows with
evolution times ${\tau}_{a,l}$ possess the involution ${\sigma}_j$
\p{auto1}, \p{auto3}, then the bosonic flows with the times $q_{l+1}$
and $t_{l+1}$ generated via formulae \p{recrel} also possess the
involution ${\sigma}_j$, but the fermionic flows with the times
${\theta}_{l+1}$ and ${\overline \theta}_{l+1}$ have
the following transformation properties with respect to ${\sigma}_j$:
\begin{eqnarray}
{\sigma}_j{\textstyle{\partial\over\partial {\theta}_{l+1}}}{\sigma}_j^{-1}
=(-1)^{l}({\textstyle{\partial\over\partial {\overline {\theta}}_{l+1}}}
+ ({\textstyle{\partial\over\partial {\overline {\theta}}_{l}}}U_0)
{\textstyle{\partial\over\partial t_1}}), \quad
{\sigma}_j{\textstyle{\partial\over\partial {\overline \theta}_{l+1}}}
{\sigma}_j^{-1} =(-1)^{l}({\textstyle{\partial\over\partial
{\theta}_{l+1}}} +({\textstyle{\partial\over\partial
{\theta}_{l}}}U_0) {\textstyle{\partial\over\partial t_1}}),
\label{transauto}
\end{eqnarray}
where we have used the following obvious relations:
${\textstyle{\partial\over\partial {\tau}_{a,l}}}H_1=
{\textstyle{\partial\over\partial q_{l}}}U_0=
{\textstyle{\partial\over\partial t_{l}}}U_0=0$. The
${\sigma}_j$-transformed flows are also admissible flows of the hierarchy,
because the additional second terms on
the right hand sides of eqs. \p{transauto} also commute 
with the $t_l$-flows\footnote{The
relations ${\textstyle{\partial\over\partial t_l}}
{\textstyle{\partial\over\partial{\tau}_{a,l}}}U_0=
{\textstyle{\partial\over\partial{\tau}_{a,l}}}
{\textstyle{\partial\over\partial t_l}}U_0=0$ are satisfied
according to the definition of the hierarchy given at the beginning
of section 2.}. So, the ${\sigma}_j$-transformation only makes a
change of basis
in the space of the fermionic flows. One can introduce a new basis
for them, which is invariant with respect to the involution
${\sigma}_j$: one defines it as the basis containing all the flows
calculated via recurrence relations \p{recrel}, except for
the ${\overline \theta}_{l}$--flows, which should be calculated via
the formula:
\begin{eqnarray}
{\textstyle{\partial\over\partial {\overline {\theta}}_{l+1}}}=
(-1)^{l}{\sigma}_j{\textstyle{\partial\over\partial
{\theta}_{l+1}}}{\sigma}_j^{-1}.
\label{transauto1}
\end{eqnarray}
In what follows we adopt such definition.

As an example, we present the second flows, i.e. those corresponding
to the evolution times $q_1$, ${\theta}_2$ and $t_2$,
\begin{eqnarray}
&&{\textstyle{\partial\over\partial q_1}}b_j =- b_j
({\alpha}_{j}\sum_{i=-\infty}^{j-1} {\overline {\alpha}}_i+ {\overline
{\alpha}}_{j} \sum_{i=j+1}^{\infty} {\alpha}_{i}), \quad
{\textstyle{\partial\over\partial q_1}}a_j =
-{\beta}_j \sum_{i=-\infty}^{j} {\overline {\alpha}}_i+
{\overline {\beta}}_{j+1} \sum_{i=j+1}^{\infty} {\alpha}_{i}, \label{flowq1}\\
&&{\textstyle{\partial\over\partial q_1}}{\beta}_j =b_{j}
\sum_{i=j}^{\infty} {\alpha}_{i}-
{\beta}_j {\overline {\alpha}}_{j} \sum_{i=j+1}^{\infty} {\alpha}_{i},
\quad {\textstyle{\partial\over\partial q_1}}{\overline {\beta}}_j=b_{j}
\sum_{i=-\infty}^{j}{\overline {\alpha}}_{i}+
{\alpha}_{j}{\overline {\beta}}_j \sum_{i=-\infty}^{j-1}
{\overline {\alpha}}_{i}, \nonumber\\
&& {\textstyle{\partial\over\partial q_1}}{\alpha}_j=
{\beta}_{j-1} + a_{j-1} \sum_{i=j}^{\infty} {\alpha}_i +
\frac{{\beta}_j{\overline {\beta}}_j}{b_{j}}
\sum_{i=j+1}^{\infty}{\alpha}_i, \quad
{\textstyle{\partial\over\partial q_1}} {\overline {\alpha}}_j=
{\overline {\beta}}_{j+1} - a_j \sum_{i=-\infty}^{j}{\overline {\alpha}}_i
-\frac{{\beta}_j{\overline {\beta}}_j}{b_{j}} \sum_{i=-\infty}^{j-1}
{\overline {\alpha}}_i, ~ ~ ~ ~ ~
\nonumber
\end{eqnarray}
\begin{eqnarray}
&&{\textstyle{\partial\over\partial {\theta}_2}}b_j =b_{j}
( -{\beta}_{j-1} -{\alpha}_{j} \sum_{i=-\infty}^{j}
(a_i+\frac{{\beta}_{i-1}{\overline {\beta}}_{i-1}}{b_{i-1}})+
(a_{j}-a_{j-1}) \sum_{i=j}^{\infty} {\alpha}_i), \label{flow13/2}\\
&& {\textstyle{\partial\over\partial {\theta}_2}}a_j = - b_{j}
{\alpha}_j-{\beta}_j \sum_{i=-\infty}^{j}
(a_i+\frac{{\beta}_{i}{\overline {\beta}}_{i}}{b_{i}})+
(b_{j+1} -b_{j} +{\beta}_j{\overline {\alpha}}_j + {\alpha}_{j+1}
{\overline {\beta}}_{j+1}) \sum_{i=j+1}^{\infty} {\alpha}_i, \nonumber\\
&&{\textstyle{\partial\over\partial {\theta}_2}}{\beta}_j =(b_{j}
{\alpha}_j - a_j {\beta}_{j}) \sum_{i=j+1}^{\infty}{\alpha}_i, \nonumber\\
&& {\textstyle{\partial\over\partial {\theta}_2}}{\overline {\beta}}_j
= -{\beta}_{j-1}{\overline {\beta}}_j + (a_j-a_{j-1}){\alpha}_j
{\overline {\beta}}_j + (a_{j-1}{\overline {\beta}}_j -
b_{j}{\overline {\alpha}}_{j}) \sum_{i=j+1}^{\infty} {\alpha}_i\nonumber\\
&&~~~~~~~~~~~~~~~~~
-(b_{j}+{\alpha}_j{\overline {\beta}}_j) \sum_{i=-\infty}^{j}
(a_i+\frac{{\beta}_{i}{\overline {\beta}}_{i}}{b_{i}}), \nonumber\\
&&{\textstyle{\partial\over\partial {\theta}_2}}{\alpha}_j={\beta}_{j-1}
\sum_{i=j}^{\infty} {\alpha}_i - {\beta}_{j} \sum_{i=j+1}^{\infty}
{\alpha}_i, \nonumber\\
&& {\textstyle{\partial\over\partial {\theta}_2}}{\overline {\alpha}}_j=
b_{j+1} - {\beta}_j {\overline {\alpha}}_j +
{\overline {\beta}}_{j+1} \sum_{i=j+2}^{\infty} {\alpha}_i -
{\overline {\beta}}_j \sum_{i=j}^{\infty}{\alpha}_i-
\frac{a_j{\beta}_j{\overline {\beta}}_j}{b_{j}}+
(a_j + \frac{{\beta}_j{\overline {\beta}}_j}{b_{j}}) \sum_{i=-\infty}^{j}
(a_i + \frac{{\beta}_i{\overline {\beta}}_i}{b_{i}}),
\nonumber
\end{eqnarray}
\begin{eqnarray}
{\textstyle{\partial\over\partial t_2}}b_j&=&b_j(a_{j}^2-a_{j-1}^2
+b_{j+1} - b_{j-1} +{\beta}_{j-1}{\overline {\alpha}}_{j-1} -
{\beta}_{j}{\overline {\alpha}}_{j}+{\alpha}_{j+1} {\overline
{\beta}}_{j+1} -{\alpha}_{j} {\overline {\beta}}_{j}),\label{flow2}\\
{\textstyle{\partial\over\partial t_2}}a_j&=&
b_{j+1} (a_{j+1}+a_{j}-{\alpha}_{j+1}{\overline {\alpha}}_{j+1}) -
b_{j} (a_{j}+a_{j-1}-{\alpha}_{j}{\overline {\alpha}}_{j})
+{\beta}_{j+1}{\overline {\beta}}_{j+1}-
{\beta}_{j}{\overline {\beta}}_{j} \nonumber \\
&+&a_j({\beta}_j{\overline {\alpha}}_j +
{\alpha}_{j+1}{\overline {\beta}}_{j+1}), \nonumber\\
{\textstyle{\partial\over\partial t_2}}{\beta}_j &=&
a_{j}^2 {\beta}_j + b_{j+1}{\beta}_{j} -b_{j}{\beta}_{j-1}+
{\beta}_j{\alpha}_{j+1}{\overline {\beta}}_{j+1}-a_{j-1} b_{j} {\alpha}_j,
\nonumber\\ {\textstyle{\partial\over\partial t_2}} {\overline {\beta}}_j
&=&-a_{j-1}^2{\overline {\beta}}_j +b_{j}{\overline {\beta}}_{j+1}-
b_{j-1}{\overline {\beta}}_{j}+{\overline {\beta}}_j {\beta}_{j-1}
{\overline {\alpha}}_{j-1} - a_{j} b_{j} {\overline {\alpha}}_{j},
\nonumber\\ {\textstyle{\partial\over\partial t_2}}{\alpha}_j&=&
b_{j} {\alpha}_{j}-b_{j-1} {\alpha}_{j-1} + a_{j}{\beta}_{j} -
a_{j-1}{\beta}_{j-1}, \nonumber\\
{\textstyle{\partial\over\partial t_2}} {\overline {\alpha}}_j
&=&b_{j+1}{\overline {\alpha}}_{j+1} -b_{j}{\overline {\alpha}}_{j} +
a_{j-1}{\overline {\beta}}_{j}-
a_{j}{\overline {\beta}}_{j+1}.
\nonumber
\end{eqnarray}
The flow with the evolution time ${\overline {\theta}}_2$ can be easily
derived from flow \p{flow13/2} via formula \p{transauto1}, and we do not
write it down here. Let us remark the nonlocality of the
${\textstyle{\partial\over\partial q_2}}$ and
${\textstyle{\partial\over\partial {\theta}_2}}$ flows.

Actually in the previous expressions \p{flowq1}, \p{flow13/2} and \p{flow2}
calculated via recurrence relations \p{recrel} we have introduced
an additional change in the space of the flows, 
\begin{eqnarray}
{\textstyle{\partial\over\partial q_1}} \rightarrow
{\textstyle{\partial\over\partial q_1}} -
S_1{\textstyle{\partial\over\partial {\overline {\theta}}_1}}+
{\overline S}_1{\textstyle{\partial\over\partial {\theta}_1}}, \quad
{\textstyle{\partial\over\partial {\theta}_2}} \rightarrow
{\textstyle{\partial\over\partial {\theta}_2}} -
H_1{\textstyle{\partial\over\partial {\theta}_1}}, \quad
{\textstyle{\partial\over\partial t_2}} \rightarrow
{\textstyle{\partial\over\partial t_2}}.
\label{newbasisf1}
\end{eqnarray}
The reason is that in this way the flows are Hamiltonian with respect to 
the second Hamiltonian structure \p{hamstr2} with Hamiltonians $U_1$, $S_2$ and 
$H_2$
\p{hams}, respectively. 

At this point one can go on and construct the Hamiltonians $U_2$,
$S_3$ and $H_3$ which generate the flows \p{flowq1}, \p{flow13/2} and
\p{flow2} via the first Hamiltonian structure \p{hamstr1}. As an example,
we present here the expression for the Hamiltonian $H_3$,
\begin{eqnarray}
H_3 = \sum_{j=-\infty}^{\infty} (\frac{1}{3} a_{j}^{3} + b_{j}
(a_j+a_{j-1}-{\alpha}_j{\overline {\alpha}}_j)
+{\beta}_j({\overline {\beta}}_j +{\overline {\beta}}_{j+1})
+a_j({\alpha}_{j+1} {\overline {\beta}}_{j+1}-
{\beta}_j{\overline {\alpha}}_j )),
\label{hamilt3}
\end{eqnarray}
which is characterized by a lattice--local Hamiltonian density.

\section{Superalgebra of the flows and reductions.}
The above constructed integrable hierarchy possesses $N=2$ supersymmetry.
Indeed, using the explicit expressions \p{flow0}--\p{flow1} for the first
four flows, one can easily check that the structure relations of the $N=2$
supersymmetry are satisfied:
\begin{eqnarray}
&& \{ {\textstyle{\partial\over\partial {\theta}_1}},
{\textstyle{\partial\over\partial {\theta}_1}} \} =
\{ {\textstyle{\partial\over\partial {\overline {\theta}}_1}},
{\textstyle{\partial\over\partial {\overline {\theta}}_1}} \} =
[ {\textstyle{\partial\over\partial t_{1}}},
{\textstyle{\partial\over\partial {\theta}_1}} ] =
[ {\textstyle{\partial\over\partial t_{1}}},
{\textstyle{\partial\over\partial {\overline {\theta}}_1}} ] =
[ {\textstyle{\partial\over\partial t_{1}}},
{\textstyle{\partial\over\partial q_0}} ] =0,\nonumber\\
&& [ {\textstyle{\partial\over\partial q_0}},
{\textstyle{\partial\over\partial {\theta}_1}} ] =
{\textstyle{\partial\over\partial {\theta}_1}}, \quad
[ {\textstyle{\partial\over\partial q_0}},
{\textstyle{\partial\over\partial {\overline {\theta}}_1}} ] =
-{\textstyle{\partial\over\partial {\overline {\theta}}_1}}, \quad
\{ {\textstyle{\partial\over\partial {\theta}_1}},
{\textstyle{\partial\over\partial {\overline {\theta}}_1}} \} =
-{\textstyle{\partial\over\partial t_{1}}},
\label{com}
\end{eqnarray}
where the bracket $\{,\}$ denotes the anticommutator. Using
flows \p{flow0}--\p{flow1} and \p{flowq1}--\p{flow2}, one can also derive
the following important (anti)commutation relations:
\begin{eqnarray}
&&\{ {\textstyle{\partial\over\partial {\theta}_1}},
{\textstyle{\partial\over\partial {\theta}_2}} \} =
\{ {\textstyle{\partial\over\partial {\theta}_2}},
{\textstyle{\partial\over\partial {\theta}_2}} \} =0,\quad
[ {\textstyle{\partial\over\partial q_0}},
{\textstyle{\partial\over\partial {\theta}_2}} ] =
{\textstyle{\partial\over\partial {\theta}_2}}, \quad
[ {\textstyle{\partial\over\partial q_1}},
{\textstyle{\partial\over\partial {\theta}_1}} ] =
{\textstyle{\partial\over\partial {\theta}_2}}, \nonumber\\
&&\{ {\textstyle{\partial\over\partial {\overline {\theta}}_1}}, 
{\textstyle{\partial\over\partial {\theta}_2}} \} =
-({\textstyle{\partial\over\partial t_{2}}}+
H_1 {\textstyle{\partial\over\partial t_{1}}}),
\label{com1}
\end{eqnarray}
where $H_1$ is the Hamiltonian from the set \p{hams}.
Let us remark that the appearance of the $H_1$ on the right hand side of
eq. \p{com1} is a basis artifact and can be excluded by introducing a
new basis in the space of Hamiltonians \p{hams}. Indeed, after introducing
the new Hamiltonians ${\widetilde U}_1$, ${\widetilde S}_2$ and
${\widetilde {\overline S}}_2$,
\begin{eqnarray}
{\widetilde U}_1=U_1+\frac{1}{2}S_1{\overline S}_1, \quad
{\widetilde S}_2=S_2-\frac{1}{2}H_1S_1,\quad
{\widetilde {\overline S}}_2={\overline S}_2-\frac{1}{2}H_1{\overline S}_1,
\label{newbasish}
\end{eqnarray}
the corresponding flows,
\begin{eqnarray}
&&{\textstyle{\partial\over\partial {\widetilde q}_1}}=
{\textstyle{\partial\over\partial q_1}} +
\frac{1}{2}S_1{\textstyle{\partial\over\partial {\overline {\theta}}_1}}-
\frac{1}{2}{\overline S}_1{\textstyle{\partial\over\partial {\theta}_1}},
\nonumber\\ && {\textstyle{\partial\over\partial {\widetilde
{\theta}}_2}}= {\textstyle{\partial\over\partial {\theta}_2}} -
\frac{1}{2}S_1{\textstyle{\partial\over\partial t_{1}}}-
\frac{1}{2}H_1{\textstyle{\partial\over\partial {\theta}_1}},  \quad
{\textstyle{\partial\over\partial {\widetilde {\overline {\theta}}}_2}}=
{\textstyle{\partial\over\partial {\overline {\theta}}_2}} -
\frac{1}{2}{\overline S}_1{\textstyle{\partial\over\partial t_{1}}}-
\frac{1}{2}H_1{\textstyle{\partial\over\partial {\overline {\theta}}_1}},
\label{newbasisf}
\end{eqnarray}
the graded commutators between the new and the remaining flows become the same
as in eqs. \p{com1}, except that the term with $H_1$ in the
last anticommutator disappears. The graded commutators
\p{com} and \p{com1} of the flows can also be derived from the Poisson
brackets \p{hamstr1}--\p{hamstr2} between the Hamiltonians
\p{def0}, \p{hams} if use is made of the homomorphism
\begin{eqnarray}
{\cal H}_{a,l}=\{{\cal H}_{b,m}, {\cal H}_{c,n}\}_2= \{{\cal H}_{b,m+1},
{\cal H}_{c,n}\}_1 \quad \Rightarrow \quad
{\textstyle{\partial\over\partial
{\tau}_{a,l}}} = [{\textstyle{\partial\over\partial {\tau}_{b,m}}},
{\textstyle{\partial\over\partial {\tau}_{c,n}}}\},
\label{homo}
\end{eqnarray}
where the brackets $[A,B\}$ between any operators $A$ and $B$ mean the
graded commutator, $[A,B\}\equiv AB-(-1)^{d_Ad_B}BA$ and it is understood 
that the pair of the indices $(a,l)$ is determined by the two pairs $(b,m)$ and 
$(c,n)$.  For example,
using \p{homo} and \p{hamsinvol}, one can derive the following
(anti)commutators:
\begin{eqnarray}
[{\textstyle{\partial\over\partial {\tau}_{a,l}}},
{\textstyle{\partial\over\partial {\tau}_{a,m}}}\}=0.
\label{homoinvol}
\end{eqnarray}

Considering the results obtained in this section, and taking into account
that the superalgebra formed by the flows should be invariant with respect
to the automorphism transformations \p{auto3}, it is reasonable to conjecture
the existence of a basis in the space of the flows where 
the algebra of the flows is:
\begin{eqnarray}
&& [{\textstyle{\partial\over\partial q_l}},
{\textstyle{\partial\over\partial q_m}}] =
\{ {\textstyle{\partial\over\partial {\theta}_l}},
{\textstyle{\partial\over\partial {\theta}_m}} \} =
\{ {\textstyle{\partial\over\partial {\overline {\theta}}_l}},
{\textstyle{\partial\over\partial {\overline {\theta}}_m}} \} =
[ {\textstyle{\partial\over\partial t_{l}}},
{\textstyle{\partial\over\partial q_m}} ] =
[ {\textstyle{\partial\over\partial t_{l}}},
{\textstyle{\partial\over\partial {\theta}_m}} ] =
[ {\textstyle{\partial\over\partial t_{l}}},
{\textstyle{\partial\over\partial {\overline {\theta}}_m}} ] =
[{\textstyle{\partial\over\partial t_l}},
{\textstyle{\partial\over\partial t_m}}] =0,\nonumber\\
&& [ {\textstyle{\partial\over\partial q_l}},
{\textstyle{\partial\over\partial {\theta}_m}} ] =
{\textstyle{\partial\over\partial {\theta}_{l+m}}}, \quad
[ {\textstyle{\partial\over\partial q_l}},
{\textstyle{\partial\over\partial {\overline {\theta}}_m}} ] =
-{\textstyle{\partial\over\partial {\overline {\theta}}_{l+m}}}, \quad
\{ {\textstyle{\partial\over\partial {\theta}_l}},
{\textstyle{\partial\over\partial {\overline {\theta}}_m}} \} =
-{\textstyle{\partial\over\partial t_{{l+m-1}}}}.
\label{generalcom}
\end{eqnarray}
 From these expressions one can see that the operator
${\textstyle{\partial\over\partial q_1}}$ acts as recursion operator
for the fermionic flows\footnote{I.e., under the adjoint action of
the operator ${\textstyle{\partial\over\partial q_1}}$ every fermionic
flow transforms into the next fermionic flow.}, while their knowledge allows
to generate the bosonic flows corresponding to the Hamiltonians $H_l$.
 
In the last part of this section we discuss the consistent reductions of 
the Toda lattice hierarchy. The three subsets of fields \p{def},
\begin{eqnarray}
&& O_1 \equiv \{ b_{i}, a_{i}, \frac{ {\beta}_{i} }{{\sqrt b_{i}}},
\frac{{\overline {\beta}}_{i} }{{\sqrt b_{i}}}, {\alpha}_{i+1},
{\overline {\alpha}}_{i} \}, \quad  m \leq i, \quad m\in {\cal N}\nonumber\\
&&O_2 \equiv \{ b_{i}, a_{i-1},a_{m-1}, {\beta}_{i},{\overline
{\beta}}_{i}, {\alpha}_{i}, {\alpha}_{m}, 
{\overline {\alpha}}_{i-1},{\overline
{\alpha}}_{m-1} \}, \quad 1 \leq i \leq m-1,
\nonumber\\ &&O_3 \equiv \{ b_{i}, a_{i-1}, \frac{ {\beta}_{i} }{{\sqrt
b_{i}}}, \frac{ {\overline {\beta}}_{i} }{{\sqrt b_{i}}}, {\alpha}_{i},
{\overline {\alpha}}_{i-1} \}, \quad i \leq 0,
\label{subsets}
\end{eqnarray}
form subalgebras of the algebras \p{hamstr1} and \p{hamstr2}.
A simple inspection of flows \p{flow0}--\p{flow2} shows that
modulo automorphism ${\sigma}_j$ \p{auto1} they admit at least two
different nonsingular reductions,
\begin{eqnarray}
I) \quad O_3=0, \quad or \quad II) \quad O_1=O_3=0.
\label{bc1}
\end{eqnarray}
Algebras \p{hamstr1}--\p{hamstr2} as well as the Hamiltonians from
the set \p{hams} and \p{hamilt3}, except $U_0$, are nonsingular on the
shell of the constraints $I)$ or $II)$ \p{bc1}. The first reduction produces
the semi-infinite lattice hierarchy where the lattice is bounded on the left
(this is related to the bosonic lattice hierarchy of the one--matrix model,
see Introduction). The
second one generates finite lattice hierarchies (the bound is on the 
left and the right simultaneously).

\section{ Origin of the $N=2$ super Toda lattice hierarchy.}
The $N=2$ discrete Toda lattice hierarchy, as it has been presented so far,
may appear to have come out of the blue.
It is time to explain how we were lead to this construction by relating it
to previously known hierarchies. Ancestors of the present paper can be
considered refs. \cite{dls,s,ls}  as well as \cite{bx}. As one might
suspect, there is a one--to--one correspondence between the lattice
hierarchy defined
above and the differential $N=2$ supersymmetric NLS hierarchy.
The relation between the two is however rather non--trivial. 
 
To start with, let us introduce a new basis  $\{ u_{j}, v_{j},
{\psi}_j, {\overline {\psi}}_{j}, {\alpha}_j, {\overline {\alpha}}_{j} \}$
in the space of the fields $\{ b_{j}, a_{j}, {\beta}_j, {\overline
{\beta}}_{j}, {\alpha}_j, {\overline {\alpha}}_{j} \}$, defined by the
following transformation:
\begin{eqnarray}
b_j = - u_j v_j, \quad a_j = (\ln
v_j)~' + \psi_j \overline \psi_j, \quad {\beta}_{j}= v_j \psi_j, \quad
{\overline {\beta}}_{j}= u_j {\overline {\psi}}_{j},
\label{transf1}
\end{eqnarray}
where the $'$ means the derivative with respect $t_1$,
and in what follows we also use the notation
${\textstyle{\partial\over\partial t_1}}\equiv {\partial}$. This
transformation is invertible, and the inverse transformation has the form:
\begin{eqnarray}
&&  u_j = - b_j {\exp (-{\partial}^{-1} (a_{j} +
\frac{{\beta}_j {\overline {\beta}}_j}{b_{j}}))}, \quad
v_j = {\exp {\partial}^{-1} (a_{j} +
\frac{{\beta}_j{\overline {\beta}}_j}{b_{j}})}, \nonumber\\ && \psi_j =
{\beta}_j {\exp (-{\partial}^{-1} (a_{j} + \frac{{\beta}_j{\overline
{\beta}}_j}{b_{j}}))}, \quad \overline \psi_j=- \frac{{\overline
{\beta}}_j}{b_j} {\exp {\partial}^{-1} (a_{j} + \frac{{\beta}_j{\overline
{\beta}}_j}{b_{j}})}.
\label{transf1inv}
\end{eqnarray}
Let us stress that the transformation \p{transf1} is nonholonomic, i.e.,
it is not reducible to a point transformation of the initial phase space.

In the new basis \p{transf1}, the first flow \p{flow1} become
\begin{eqnarray}
&&(\ln(u_{j+1}v_j))~'=\psi_{j+1}\overline \psi_{j+1}-\psi_j \overline
\psi_j, \quad
(\ln v_j)~''= u_{j}v_{j}-u_{j+1}v_{j+1} +
\psi_{j+1}~'~\overline \psi_{j+1}-\psi_j~'~\overline \psi_j, \nonumber\\
&& ({ \psi_{j}~'\over u_{j}})~'=v_{j}\psi_{j}-v_{j-1}\psi_{j-1}, \quad
({\overline \psi_j~'\over v_j})~'=
u_j \overline \psi_j -u_{j+1}\overline \psi_{j+1}, \quad
\psi_{j}~'=u_{j}{\alpha}_j, \quad
\overline \psi_j~'=v_j{\overline {\alpha}}_j,{}~~~
\label{ftoda}
\end{eqnarray}
and coincides with the  
minimal $N=2$ supersymmetric Toda chain
equations, introduced in \cite{dls,ls} and called f--Toda,  
extended by two
additional fields ${\alpha}_j$ and $ {\overline {\alpha}}_j$. They are
compatible with the fermionic flows
\begin{eqnarray}
&&{\textstyle{\partial\over\partial {\theta}_1}}u_j=-{\psi}_{j}~',\quad
{\textstyle{\partial\over\partial {\theta}_1}}v_j=0,\quad
{\textstyle{\partial\over\partial {\theta}_1}}{\psi}_j =0, \quad
{\textstyle{\partial\over\partial {\theta}_1}}{\overline {\psi}}_j=v_{j},
\quad {\textstyle{\partial\over\partial {\theta}_1}}{\alpha}_j=0,\quad
{\textstyle{\partial\over\partial {\theta}_1}}
{\overline {\alpha}}_j= (\ln v_{j})~',{}~~~~
\label{nflow11/2}
\end{eqnarray}
\begin{eqnarray}
&&{\textstyle{\partial\over\partial {\overline {\theta}}_1}}u_j =0, \quad
{\textstyle{\partial\over\partial {\overline {\theta}}_1}}v_{j}=
-{\overline {\psi}}_{j}~', \quad
{\textstyle{\partial\over\partial {\overline {\theta}}_1}}{\psi}_j=u_{j},
\quad {\textstyle{\partial\over\partial {\overline {\theta}}_1}}
{\overline {\psi}}_j=0, \quad
{\textstyle{\partial\over\partial {\overline {\theta}}_1}}{\alpha}_j
=(\ln u_{j})~', \quad
{\textstyle{\partial\over\partial {\overline {\theta}}_1}}
{\overline {\alpha}}_j=0.{}~~~~ 
\label{nflow21/2}
\end{eqnarray}
These correspond to vector fields that generate the transformations of 
the $N=2$ supersymmetry. 

In \cite{dls} it was demonstrated that the flow equations
belonging to the differential $N=2$ NLS  
hierarchy, are invariant with respect to the f--Toda transformations in the
sense that different solutions are related by f--Toda transformations.
Let us consider an example involving the second flow equations of the $N=2$ 
NLS hierarchy \cite{rk,kst}. These equations are 
\begin{eqnarray}
&& {\textstyle{\partial\over\partial t_2}}u_j=
-u_j~''+2u_{j}^2v_j + 2u_j({\psi}_j{\overline \psi}_j~'-
{\psi}_j~'{\overline \psi}_j)+ 2u_j~'{\psi}_j{\overline \psi}_j, \nonumber\\
&&{\textstyle{\partial\over\partial t_2}}v_j =
v_j~''-2u_jv_j^2 - 2v_j({\psi}_j{\overline \psi}_j~'-
{\psi}_j~'~{\overline \psi}_j)+2v_j~'{\psi}_j{\overline \psi}_j,\nonumber\\
&& {\textstyle{\partial\over\partial t_2}} {\psi}_j= -{\psi}_j~''+
2u_jv_j{\psi}_j-2{\psi}_j~'~{\overline \psi}_j {\psi}_j,
\quad {\textstyle{\partial\over\partial t_2}}
{\overline \psi}_j= {\overline \psi}_j~''- 2u_jv_j{\overline \psi}_j-
2{\psi}_j {\overline \psi}_j~'~ {\overline \psi}_j,
\label{nls}
\end{eqnarray}
where here we have appended a label $j$ in order to mark different solutions. 
Solutions labeled by neighboring $j$'s are related by the f--Toda chain 
equations \p{ftoda}. In fact, using
the transformations \p{transf1} and \p{transf1inv}, one can rewrite eqs.
\p{nls} as well as eqs. \p{nflow11/2} and \p{nflow21/2}
in terms of the fields $ b_{j}, a_{j}, {\beta}_j,
{\overline {\beta}}_{j}$. Thus, in the old basis they become
\begin{eqnarray}
&& {\textstyle{\partial\over\partial t_2}}b_j=-b_j~''+2(b_{j}a_j)~' +
2b_j(\frac{{\beta}_j{\overline {\beta}}_j}{b_{j}})~',
\nonumber\\ &&{\textstyle{\partial\over\partial t_2}}a_j = a_j~''+
2a_j~'a_j + 2b_j~' +2({\beta}_j(\frac{{\overline {\beta}}_j}{b_{j}})~'+
a_i\frac{{\beta}_j{\overline {\beta}}_j}{b_{j}})~',
\nonumber\\ && {\textstyle{\partial\over\partial t_2}} {\beta}_j= -
{\beta}_j~''+ 2(a_j{\beta}_j)~', \quad {\textstyle{\partial\over\partial
t_2}} {\overline \beta}_j= {\overline \beta}_j~''+ 2(a_j{\overline
\beta}_j)~'- 2({\overline \beta}_j ({\ln b_j})~'~)~',
\label{nkdv}
\end{eqnarray}
\begin{eqnarray}
&&{\textstyle{\partial\over\partial {\theta}_1}}b_j=
{\beta}_{j}~'-a_j {\beta}_{j},\quad
{\textstyle{\partial\over\partial {\theta}_1}}a_j=-{\beta}_j,\nonumber\\
&& {\textstyle{\partial\over\partial {\theta}_1}}{\beta}_j =0, \quad
{\textstyle{\partial\over\partial {\theta}_1}}{\overline {\beta}}_j
=-b_{j}-a_j \frac{{\beta}_j{\overline {\beta}}_j}{b_{j}} +
\frac{{\beta}_j~'~{\overline {\beta}}_j}{b_{j}}, \nonumber\\
&&{\textstyle{\partial\over\partial {\theta}_1}}{\alpha}_j=0,\quad
{\textstyle{\partial\over\partial {\theta}_1}}
{\overline {\alpha}}_j= a_{j} + \frac{{\beta}_j{\overline
{\beta}}_j}{b_{j}},
\label{nnflow11/2}
\end{eqnarray} \begin{eqnarray}
&&{\textstyle{\partial\over\partial {\overline {\theta}}_1}}b_j =
a_j{\overline {\beta}}_{j} + b_j (\frac{{\overline {\beta}}_j}{b_{j}})~',
\quad {\textstyle{\partial\over\partial {\overline {\theta}}_1}}a_{j}=
{\overline {\beta}}_{j} + (\frac{{\overline {\beta}}_j}{b_{j}})~''+
(\frac{a_j{\overline {\beta}}_j}{b_{j}})~', \nonumber\\
&& {\textstyle{\partial\over\partial {\overline {\theta}}_1}}{\beta}_j
=-b_{j}-{\beta}_j (\frac{{\overline {\beta}}_j}{b_{j}})~'-
a_j\frac{{\beta}_j{\overline {\beta}}_j}{b_{j}},\quad
{\textstyle{\partial\over\partial {\overline {\theta}}_1}}
{\overline {\beta}}_j=0, \nonumber\\
&&{\textstyle{\partial\over\partial {\overline {\theta}}_1}}{\alpha}_j
=-a_{j} + (\ln b_j)~' - \frac{{\beta}_j{\overline {\beta}}_j}{b_{j}}, \quad
{\textstyle{\partial\over\partial {\overline {\theta}}_1}}
{\overline {\alpha}}_j=0,
\label{nnflow21/2}
\end{eqnarray}
respectively. One can check easily by direct substitution of the flows
\p{flow11/2}, \p{flow21/2}, \p{flow1} and \p{flow2} into eqs. \p{nkdv},
\p{nnflow11/2} and \p{nnflow21/2} that they are indeed identically
satisfied. Of course, the same procedure can be applied to any flow
belonging to the $N=2$ supersymmetric NLS hierarchy. Thus,
flows of the $N=2$ supersymmetric Toda lattice hierarchy can be
reconstructed using the set of the corresponding $N=2$ supersymmetric NLS
flows via the f--Toda chain equations. 

As one can expect at this point, the recursion
operators of the $N=2$ supersymmetric Toda lattice and NLS hierarchies
are related. Indeed, if we use the first flow equations \p{flow1} we can
derive the following relations:
\begin{eqnarray}
&& a_{j-1}=a_{j}- ({\ln b_j})~', \quad
b_{j+1} + {\alpha}_{j+1}{\overline {\beta}}_{j+1}=
a_j~' + b_{j} - {\beta}_j{\overline {\alpha}}_j, \nonumber\\
&& {\beta}_{j-1}={\beta}_j - {\alpha}_j~', \quad
{\overline {\beta}}_{j+1}={\overline {\beta}}_j-
{\overline {\alpha}}_j~', \quad
\sum_{j=-\infty}^{i} {\alpha}_j={\partial}^{-1}{\beta}_i, \quad
\sum_{j=i}^{\infty} {\overline \alpha}_j={\partial}^{-1}{\overline \beta}_i,
\nonumber\\ &&\sum_{j=i+1}^{\infty}{\ln b_j}=-{\partial}^{-1}a_i, \quad
\sum_{j=-\infty}^{i-1}( a_{j} +\frac{{\beta}_j{\overline
{\beta}}_j}{b_{j}})={\partial}^{-1}(b_i+{\alpha}_{i}{\overline
{\beta}}_{i}), 
\label{flow1rel}
\end{eqnarray}
which, upon substitution into the recurrence relations \p{recrel} transform
them into relations involving only the fields $\{ b_{i}, a_{i},
{\beta}_{i}, {\overline {\beta}}_{i}, {\alpha}_{i}, {\overline
{\alpha}}_{i} \}$ defined at the same lattice point. After applying
transformations \p{transf1}--\p{transf1inv} to the new basis $\{ u_{j},
v_{j}, {\psi}_j, {\overline {\psi}}_{j}, {\alpha}_j= {\psi_{j}}~'/u_{j},
{\overline {\alpha}}_{j} = {\overline \psi_j}~'/v_j \}$, one can verify
that the resulting relations reproduce the corresponding recursion
relations for the $N=2$ supersymmetric NLS hierarchy derived in \cite{dls}
in terms of $N=2$ superfields. For the bosonic Toda lattice
hierarchy a
similar procedure was developed in \cite{bx}.

We would also like to note that at least three sets of Hamiltonians
\p{def0}, $S_l$, ${\overline S}_l$ and $H_l$, can be recovered using the
connection between the $N=2$ supersymmetric Toda lattice and NLS
hierarchies. Let us shortly explain the main steps of such procedure.

We take the sum of the f--Toda chain equations \p{ftoda} over the
lattice points and differentiate with respect to some
evolution time ${\cal T}_l$ of the $N=2$ super NLS hierarchy.
We remark that the right--hand sides of the resulting
expressions become identically equal to zero and the left--hand sides
become the full derivatives with respect to the time $t_1$ corresponding
to the first flow of the $N=2$ supersymmetric Toda lattice hierarchy,
\begin{eqnarray}
&& {\textstyle{\partial\over\partial t_1}}
[\sum_{j=-\infty}^{\infty}{\textstyle{\partial\over\partial {\cal T}_l}}
(\ln(u_{j+1}v_j))]=0, \quad
{\textstyle{\partial\over\partial t_1}}[\sum_{j=-\infty}^{\infty}
{\textstyle{\partial\over\partial {\cal T}_l}}({ \psi_{j}~'\over u_{j}})]
=0, \quad {\textstyle{\partial\over\partial t_1}}
[\sum_{j=-\infty}^{\infty}{\textstyle{\partial\over\partial {\cal T}_l}}
({\overline \psi_j~'\over v_j})]=0,
\label{ftodaint}
\end{eqnarray}
where we have used a very important property, the commutativity
the $N=2$ super NLS and f--Toda flows,
$[{\textstyle{\partial\over\partial t_1}},
{\textstyle{\partial\over\partial {\cal T}_l}}]$=0.
The equations \p{ftodaint} have the form of conservation laws, and
it is obvious that the expressions inside the square brackets are the
Toda--lattice Hamiltonians\footnote{In eqs. \p{ftodaint}
only the relations generating nontrivial independent
Hamiltonians are presented.}. Thus, any flow of the $N=2$ super
NLS hierarchy can generate some Hamiltonian for the $N=2$ supersymmetric
Toda lattice hierarchy. Now we substitute the explicit expressions
for the
first three bosonic and fermionic flows of the $N=2$ super NLS
hierarchy into the expressions inside the square brackets of eqs.
\p{ftodaint}; then we use eqs.\p{ftoda} in order to express higher
derivatives with respect to $t_1$ that appear in the calculations
in terms of its first and zeroth derivatives; we make the transformations
\p{transf1}--\p{transf1inv} and eliminate the operator ${\partial}^{-1}$
via formulae \p{flow1rel}. We have checked that the expressions
inside the square brackets of \p{ftodaint} indeed reproduce
the integrals $S_1,S_2,{\overline S}_1,{\overline S}_2, H_1, H_2, H_3$
\p{hams}, \p{hamilt3}.

The above illustrated connection with the $N=2$ NLS hierarchy via the f--Toda
equations explains the origin of the definitions and properties of the
previous sections. Let us end this section with two remarks.

First, we can extend the above correspondence to the $a=4$ $N=2$ 
super-KdV \cite{lm} hierarchy. 
Using the mapping \cite{kst} which connects the $N=2$ super-NLS and the
$a=4$ $N=2$ super-KdV hierarchies, as well as using
transformations \p{transf1} and \p{transf1inv}, one can derive the mapping
\begin{eqnarray}
&& s_j = - a_j, \quad r_j = b_{j}+{\beta}_j ((\frac{{\overline
{\beta}}_j}{b_{j}})~'+ a_i\frac{{\overline {\beta}}_j}{b_{j}}), \nonumber\\
&&\xi_j = \frac{1}{2} {\beta}_j, \quad \overline \xi_j=-\frac{1}{2}
{\overline {\beta}}_j - \frac{1}{2} ((\frac{{\overline {\beta}}_j}{b_{j}})~'
+ a_i\frac{{\overline {\beta}}_j}{b_{j}})~', 
\label{transf2}
\end{eqnarray}
which connects our eqs. \p{nkdv} and the second flow equations
\begin{eqnarray}
&& {\textstyle{\partial\over\partial t_2}}r_j=-r_j~''-2(r_{j}s_j)~' -
8({\xi}_j{\overline {\xi}}_j)~', \quad
{\textstyle{\partial\over\partial t_2}}s_j = s_j~''-2s_j~'s_j - 2r_j~',
\nonumber\\
&& {\textstyle{\partial\over\partial t_2}} {\xi}_j= - {\xi}_j~''-
2(s_j{\xi}_j)~', \quad
{\textstyle{\partial\over\partial t_2}}
{\overline \xi}_j= {\overline \xi}_j~''- 2(s_j{\overline \xi}_j)~'
\label{kdv}
\end{eqnarray}
belonging to the $a=4$ $N=2$ super-KdV hierarchy. Here $s_j$ and $r_j$
($\xi_j$ and ${\overline \xi}_j$) are bosonic (fermionic) fields with
the scaling dimensions $1$ and $2$ ($3/2$ and $3/2$), respectively.

Finally we would like to point out the possibility
to apply the approach developed in this section to the wide class of the
$N=2$ supersymmetric generalized Toda lattice equations constructed
recently in \cite{s}. They are related with the $N=2$ supersymmetric
$(n,m)$ Generalized NLS hierarchies (GNLS) \cite{bks} in the same way as
the above $N=2$ supersymmetric Toda lattice and NLS hierarchies are
related. Recently the recursion operators for the $N=2$
super $(n,m)$--GNLS hierarchies were constructed in \cite{bs}. Using
them and the above method one can derive  the recursion
operators for the corresponding $N=2$ supersymmetric generalized
Toda lattice hierarchies. The details will be given elsewhere.

\section{Lax--pair representation of the $N=2$ super
Toda lattice hierarchy.}
The more compact way to formulate an integrable hierarchy is 
by means of a Lax--pair. In the present case we can do that
by means of the recursion operator $R_{AB,ij}$ \p{supmatr}--\p{recopb2}.
In terms of it
the Lax--pair representation for the bosonic flows of the $N=2$
supersymmetric Toda lattice hierarchy is:
\begin{eqnarray}
{\textstyle{\partial\over\partial {\tau}_{a,l}}} R_{AB,ij} =
\sum_{k=-\infty}^{\infty}\sum_{C=1}^{6}
((K~^{'}_{a,l})_{AC,ik} R_{CB,kj} - R_{AC,ik}
(K~^{'}_{a,l})_{CB,kj}), \quad a=1,4,
\label{Lax}
\end{eqnarray}
where $(K~^{'}_{a,l})_{AC,ik}$ is the matrix of Fr\'echet
derivatives of the function $(K_{a,l})_{A,i}$ defined in eq.
\p{flows}.

For completeness we present also the operator representation for the
fermionic flows,
\begin{eqnarray}
{\epsilon} {\textstyle{\partial\over\partial {\tau}_{a,l}}}
R_{AB,ij} =
\sum_{k=-\infty}^{\infty}\sum_{C=1}^{6}
({\epsilon}(K~^{'}_{a,l})_{AC,ik} R_{CB,kj} + R_{AC,ik}{\epsilon}
(K~^{'}_{a,l})_{CB,kj}), \quad a=2,3,
\label{Laxf}
\end{eqnarray}
where we have introduced an additional Grassmann parameter ${\epsilon}$ in order
to derive a closed relation containing only two supermatrices,
$R_{AB,ij}$ and $(K~^{'}_{a,l})_{AC,ik}$ at $a=2,3$.

The representation \p{Lax} can be treated as the integrability condition
for the linear system
\begin{eqnarray}
&&\sum_{j=-\infty}^{\infty}\sum_{C=1}^{6}R_{AC,ij}{\cal N}_{C,j} =
{\lambda} {\cal N}_{A,i}, \nonumber\\
&& {\textstyle{\partial\over\partial {\tau}_{a,l}}} {\cal N}_{A,i} =
\sum_{j=-\infty}^{\infty}\sum_{C=1}^{6} (K~^{'}_{a,l})_{AC,ij}
{\cal N}_{C,j}, \quad a=1,4,
\label{linsys}
\end{eqnarray}
where ${\cal N}_{C,j}$ are its eigenfunctions and $\lambda$ is the
spectral parameter. The Hamiltonians which are in involution can be
derived using the formula\footnote{Let us recall the definition of the
supertrace for a supermatrix of the form \p{supmatr}: $str(R)=
tr(B_{(2\infty) \times (2\infty)})-tr(B_{(4\infty) \times (4\infty)})$.}
\begin{eqnarray}
H_l= str(R^l).
\label{hams1}
\end{eqnarray}
Acting $p$-times with the hereditary recursion operator
\p{supmatr}--\p{recopb2} on the first Hamiltonian structure $J_1$
\p{hamstr1} one can derive the $(p+1)$-th Hamiltonian structure,
\begin{eqnarray}
J_{p+1}=R^p J_1,
\label{hamstrn}
\end{eqnarray}
which is compatible$^1$ with $J_1$. Thus, almost all information about the
whole hierarchy is encoded in the recursion operator
\p{supmatr}--\p{recopb2}.

Let us remark the existence of another linear system which includes
only bosonic wave functions. One can check by a direct computation
that the following linear system
\begin{eqnarray}
&& (1 + \frac{{\beta}_{j+1}
(a_{j}{\overline {\beta}}_j + b_{j}{\overline {\alpha}}_j +
{\alpha}_{j+1}{\overline {\alpha}}_j {\overline {\beta}}_j)}{b_{j}b_{j+1}})
{\cal M}_{j+1} +
(b_j + \frac{(a_j (b_{j-1}-b_j) {\beta}_j
+ b_j^2 {\alpha}_j){\overline {\beta}}_{j-1} }{b_{j-1}b_{j}})
{\cal M}_{j-1} \nonumber\\
&& + (a_j + \frac{{\beta}_{j}{\overline {\beta}}_{j-1}}{b_{j-1}} +
{\alpha}_{j+1}{\overline {\alpha}}_j -
\frac{a_j{\alpha}_{j+1}{\overline {\beta}}_{j}}{b_{j}}+
\frac{a_{j+1}(b_{j+1}-b_j) {\beta}_{j+1}{\overline {\beta}}_j
{\alpha}_{j+1}{\overline {\alpha}}_j}{b_{j}b_{j+1}^2}){\cal M}_j = 0,
\label{laxweekspectr}
\end{eqnarray}
\begin{eqnarray}
{\textstyle{\partial\over\partial t_1}}{\cal M}_j=-
(b_j + \frac{(a_j (b_{j-1}-b_j) {\beta}_j
+ b_j^2 {\alpha}_j){\overline {\beta}}_{j-1} }{b_{j-1}b_{j}})
{\cal M}_{j-1} + {\beta}_{j} (\frac{{\overline {\beta}}_j}{b_{j}} -
\frac{{\overline {\beta}}_{j-1}}{b_{j-1}}) {\cal M}_j
\label{laxweek}
\end{eqnarray}
is consistent if the first flow equations \p{flow1} are satisfied.
Here, ${\cal M}_j$ are bosonic wave functions which in consequence of
\p{flow1} and \p{laxweekspectr}-\p{laxweek} satisfy the following second
order differential equation
\begin{eqnarray}
{\textstyle{{\partial}^2\over\partial t_1^2}}{\cal M}_j-
(a_{j} + \frac{{\beta}_j{\overline {\beta}}_j}{b_{j}})
{\textstyle{\partial\over\partial t_1}}{\cal M}_j +
(b_j + {\beta}_j{\overline {\alpha}}_j + {\alpha}_{j}
{\overline {\beta}}_{j}+ \frac{a_j {\beta}_j
{\overline {\beta}}_{j}}{b_{j}}) {\cal M}_j =0
\label{laxweekeq}
\end{eqnarray}
for any values of the index $j$. However, the linear system
\p{laxweekspectr}--\p{laxweek} provides a Lax--pair representation 
valid only in a week sense, i.e. only when the operator equation 
are restricted on
the shell of the wave functions satisfying eq. \p{laxweekspectr}. Unfortunately
this is an obstacle at least for a straightforward derivation of the higher
Hamiltonians and flows.

\section{Bosonic limit.}
In the previous section we found two Lax pair representation of our
hierarchy, although the second one is only a `conditional' Lax pair.
To give a closer look to such Lax pairs, it is convenient to
examine their bosonic limit.  
In this limit (i.e., as ${\alpha}_{j}={\overline {\alpha}}_j=
{\beta}_{j}={\overline {\beta}}_j=0$) it is possible to introduce a
spectral parameter into the linear system \p{laxweekspectr}--\p{laxweek} to
derive the Lax--pair representation in a
strong, operator form. Indeed, in the bosonic
limit the first flow \p{flow1} admit the one parameter group of 
invariance transformations $a_j \rightarrow a_j -\gamma$, where $\gamma$
is an arbitrary constant parameter. Applying this transformation to
the bosonic limit of the linear system \p{laxweekspectr}-\p{laxweek}, 
it becomes
\begin{eqnarray}
&& (L_0 {\cal M})_i \equiv {\cal M}_{j+1} +b_j {\cal M}_{j-1} +
a_j {\cal M}_j = {\gamma} {\cal M}_j, \nonumber\\
&&{\textstyle{\partial\over\partial t_1}}{\cal M}_j=-b_j{\cal M}_{j-1}
\equiv (A_0 {\cal M})_i
\label{laxlimstr}
\end{eqnarray}
and reproduces the well known form for the linear system corresponding
to the bosonic Toda lattice (see, e.g., \cite{ft} and references
therein). This means that the Toda lattice equations are 
the integrability conditions of the Lax--pair representation,
\begin{eqnarray}
{\textstyle{\partial\over\partial t_1}} L_0 =[A_0,L_0],
\label{laxlimrepr}
\end{eqnarray}
which is valid in a strong, operator sense. This Lax pair is to be 
compared with the one in the Introduction, which appears in the one--matrix
model. 

It is
interesting to remark that in \p{laxlimstr} the transformation parameter
$\gamma$ plays the role of the spectral parameter. The supersymmetric flow
\p{flow1} does not admit such transformation, and, actually, it is not
clear how to introduce a spectral parameter into \p{laxweekspectr} (which
would be necessary in order to construct a Lax--pair representation in a strong,
operator sense). Moreover, in the case under consideration, it is even not
obvious that it is possible at all, because the linear system includes
only bosonic wave functions\footnote{Let us recall that linear systems
corresponding to supersymmetric integrable equations contain, as a rule,
both bosonic and fermionic wave functions.}.

Let us come now to a comparison betwen the bosonic limit of the Lax--pair
representation \p{Lax}, \p{linsys} with the representation
\p{laxlimstr}-\p{laxlimrepr}. In the bosonic limit the
recursion operator \p{supmatr}--\p{recopb2} has a block-diagonal structure
and can be represented by a direct sum of the following two matrices:
\begin{eqnarray}
&& R_1= \pmatrix{a_{i-1} +
\frac{b_{i}}{b_{j}} ( a_{i-1} - a_{i}) \sum_{k \geq1} \delta_{i,j-k}, &
b_{i} (\delta_{i,j} + \delta_{i,j+1}) \cr -\frac{1}{b_{j}} (b_{i+1}
\sum_{k \geq 1} \delta_{i,j-k-1} - b_{i} \sum_{k \geq 1 }
\delta_{i,j-k+1}), & a_i \delta_{i,j} \cr}, \nonumber\\ && R_2=
\pmatrix{0,&0, & -b_{i} \sum_{k \geq 1} \delta_{i,j+k},&0\cr 0,&0, &0,&
b_{i} \sum_{k \geq 1} \delta_{i,j-k} \cr \delta_{i,j+1}, & 0,&
-a_{i-1}\sum_{k \geq 1}\delta_{i,j+k},&0\cr 0,&-\delta_{i,j-1}, & 0,&
-a_{i}\sum_{k \geq 1}\delta_{i,j-k}\cr}.
\label{boslax1}
\end{eqnarray}
The matrix $R_1$ is the recursion operator of the bosonic Toda lattice
hierarchy, while the interpretation of the matrix $R_2$ is a bit
obscure. For it is not easy to interpret the following fact: the
traces
$tr(R^{l}_2)$ are equal to zero at least for the first two values of the
$l=1$ and $l=2$, and it is not possible to generate the Toda lattice
Hamiltonians $h_l$ using the standard prescription $h_l=tr(R^{l}_2$).

Calculating the matrix of Fr\'echet derivatives 
$(K~^{'}_{4,1})_{AC,ij}$ 
corresponding to the first flow \p{flow1} and finding its bosonic
limit\footnote{One should pay attention to the fact that these operations are
not commutative, so, their order is crucial.}, one can observe that it
also splits into a direct sum of two matrices,
\begin{eqnarray}
&& K~^{'}_1= \pmatrix{( a_{i} - a_{i-1}) \delta_{i,j}, &
b_{i} (\delta_{i,j} - \delta_{i,j+1}) \cr
\delta_{i,j-1} - \delta_{i,j}, & 0 \cr}, \nonumber\\
&& K~^{'}_2= \pmatrix{a_i\delta_{i,j},&0, & -b_{i} \delta_{i,j},&0\cr
0,&-a_{i-1}\delta_{i,j}, &0,& -b_{i} \delta_{i,j} \cr
\delta_{i,j} - \delta_{i,j+1},&0,& 0,& 0\cr
0,&\delta_{i,j} - \delta_{i,j-1},&0,& 0\cr},
\label{boslax2}
\end{eqnarray}
respectively. Comparison between \p{boslax1}-\p{boslax2} and \p{laxlimstr}
shows that in this way we produce two different Lax--pair
representations with the matrices $R_1,K~^{'}_1$ and $R_2,K~^{'}_2$,
respectively, which do not coincide with the corresponding matrices
$L_0,A_0$ of the representation \p{laxlimstr}. Moreover, their algebraic
origins are also different: the dimension of the Lax operator $L_0$
\p{laxlimstr} corresponds to the fundamental representation of the
$sl(\infty)$ algebra, while the dimension of the recursion operator $R_1$
\p{boslax1} corresponds to its adjoint representation. The
former representation has a simpler matrix structure, and, in this
sense, it is preferable to the latter. Moreover the
representation \p{laxlimstr} arises naturally in the context of the bosonic 
one-matrix model, \cite{bmx}. It is
plausible to think that it is also relevant for the corresponding
supermatrix model, if any. However the relationship between the two
representations is not evident, and, for the time being, we can say that 
they are complementary representations.  

Let us comment now on the general structure of
the supersymmetric counterpart of the first representation, i.e., 
supersymmetric Lax--pair
representation whose bosonic limit contains the Lax operator
$L_0$ \p{laxlimstr}. Keeping in mind the analogy with the bosonic case, it
is reasonable to conjecture that a general structure of such supersymmetric
Lax operator $L$ is similar to the general structure \p{supmatr} of the
recursion operator $R$, but their dimensions are different, and
the relation between them is the same as in the bosonic case. Finally, we
conjecture the following form of $L$ (compare \p{supmatrlax} with \p{supmatr}):
\begin{eqnarray}
L=\pmatrix{B_{(\infty) \times (\infty)}, &
F_{(\infty) \times 2(\infty)}\cr F_{(2\infty) \times (\infty)}, &
B_{(2\infty) \times (2\infty)} \cr}, \quad
B_{(\infty) \times (\infty)}|=L_0,
\label{supmatrlax}
\end{eqnarray}
where $|$ denotes the bosonic limit. The matrix \p{supmatrlax} corresponds
to the fundamental representation of the $sl(\infty|2\infty)$
superalgebra. As for the matrix $B_{(2\infty) \times (2\infty)}|$,
it either corresponds to some Lax operator of the bosonic Toda lattice or
identically vanishes.

\section{ Conclusion.}
In this paper we have presented in any detail a new $N=2$ supersymmetric Toda 
lattice hierarchy and indicated the method to construct generalizations of
it. The former can be thought of as the discrete version of the $N=2$
differential NLS hierarchy. The latter would correspond to the discrete 
version of the differential $N=2$ GNLS hierarchies, \cite{bks}.
A few years ago $N=2$ discrete Toda lattice hierarchies were proposed,
\cite{i}: it would be interesting to know whether they bear any relation
to our hierarchies. The relation, if it exists, is rather non--trivial.
There appear to be no straightforward restriction that may lead to our 
hierarchies, perhaps some more complicated coset construction is necessary. 

Finally we would like to make some comments on relevant consequences of our 
result. The $N=2$ hierarchy presented in this paper, as pointed out
in the introduction, is the extension of the discrete integrable hierarchy that
arises in the one--matrix random model. Starting from the Virasoro 
constraints of the latter and the fact that in any model these 
constraints must be consistent with the underlying hierarchy, we are very 
likely to already possess all the information we need in order to write down 
the complete set of super--Virasoro constraints consistent with the $N=2$
Toda lattice hierarchy. This means that we can completely calculate
the free energy and correlators of the would--be $N=2$ one--matrix model
(see, for example, the second reference of \cite{bx})
and, therefore, to completely determine it. Moreover, the correlators
in the non-supersymmetric case are interpretable as correlators
of a topological field theory. Therefore, once we know the correlators
in the $N=2$ case, it will be interesting to examine what kind of new 
information about topological field theories they may provide.

\vskip1cm

\noindent{\bf Acknowledgments.}
We would like to thank  L. Alvarez-Gaume, E. Ivanov, I. Krichever,
D. Lebedev, G. Marmo, A. Morozov, V. Rubtsov and especially O. Lechtenfeld
and A. Leznov for many useful and clarifying discussions. This work was
partially supported by the Russian Foundation for Basic Research, Grant
No. 96-02-17634, RFBR-DFG Grant No 96-02-00180, INTAS Grant No. 93-1038,
INTAS Grant No. 93-127, INTAS Grant No. 94-2317, a grant from the Dutch
NWO organization and by the EC TMR Programme, grant FMRX-CT96-0012.

\section*{Appendix. Canonical basis and 
Lagrangian formulation of the $t_1$-flow equations.}
It is interesting that the first flow \p{flow1} of our $N=2$
hierarchy admits
a Lagrangian formulation. Let us introduce the new coordinates
$\{ x_{j}, p_{j}, {\xi}_j, {\overline {\xi}}_{j},
{\eta}_j, {\overline {\eta}}_{j} \}$ in the phase space \p{def}
\begin{eqnarray}
&&b_j =  e^{x_j-x_{j-1}}, \quad a_j = -p_j, \nonumber\\
&&{\beta}_{j}=- e^{x_j} \xi_j, \quad {\overline {\beta}}_{j}=
e^{-x_{j-1}}{\overline {\xi}}_{j}, \nonumber\\
&&{\alpha}_j= {\eta}_{j-1} - {\eta}_{j}, \quad
{\overline \alpha}_j= {\overline \eta}_{j}
\label{transcan1}
\end{eqnarray}
for which the first Hamiltonian structure \p{hamstr1} becomes canonical,
\begin{eqnarray}
\{ x_i, p_j \}_1=\delta_{i,j}, \quad
\{ {\xi}_i,{\overline {\xi}}_j \}_1= \delta_{i,j}, \quad
\{ {\eta}_i,{\overline {\eta}}_j \}_1= \delta_{i,j}.
\label{hamstrcan1}
\end{eqnarray}
In terms of these coordinates the Hamiltonian $H_2$ \p{hams} is 
\begin{eqnarray}
H_2=\sum_{j=-\infty}^{\infty} (\frac{1}{2} p_{j}^{2}+e^{x_j-x_{j-1}}+
e^{x_j} \xi_j{\overline \eta}_{j}+e^{-x_{j-1}}({\eta}_{j-1} -
{\eta}_{j}){\overline {\xi}}_{j})
\label{ham2can}
\end{eqnarray}
It generates the following $t_1$-flow equations \p{flow1} via the first 
Hamiltonian structure
\footnote{Here, for convenience we have
reversed the $t_1$-sign, $t_1\rightarrow -t_1$.},
\begin{eqnarray}
&&{\textstyle{\partial\over\partial t_1}}x_j=p_{j}, \quad
{\textstyle{\partial\over\partial t_1}}p_j=
e^{x_{j+1}-x_{j}}-
e^{x_j-x_{j-1}}-
e^{x_j}{\xi}_j{\overline {\eta}}_j +
e^{-x_{j}}({\eta}_{j}-{\eta}_{j+1}){\overline {\xi}}_{j+1},\nonumber\\
&& {\textstyle{\partial\over\partial t_1}}{\xi}_j
=e^{-x_{j-1}}({\eta}_{j}-{\eta}_{j-1}), \quad
{\textstyle{\partial\over\partial t_1}}{\overline {\xi}}_j
=e^{x_j}{\overline {\eta}}_j, \nonumber\\
&&{\textstyle{\partial\over\partial t_1}}{\eta}_j
=-e^{x_j}{\xi}_j, \quad
{\textstyle{\partial\over\partial t_1}}{\overline {\eta}}_j
=e^{-x_{j}}{\overline {\xi}}_{j+1}-e^{-x_{j-1}}{\overline {\xi}}_{j},
\label{flow1can}
\end{eqnarray}
admitting the automorphism ${\sigma}_j$ \p{auto1}, which has the following
realization in terms of the new coordinates \p{transcan1}:
\begin{eqnarray}
&& \sigma_j x_i {\sigma}^{-1}_j= -x_{j-i-1}, \quad
\sigma_j p_i {\sigma}^{-1}_j=-p_{j-i-1} , \nonumber\\
&& \sigma_j {\xi}_i {\sigma}^{-1}_j=-{\overline {\xi}}_{j-i}, \quad
\sigma_j {\overline {\xi}}_i {\sigma}^{-1}_j=- {\xi}_{j-i}, \nonumber\\
&& \sigma_j {\eta}_i {\sigma}^{-1}_j=-\sum_{k=-i}^{\infty}
{\overline {\eta}}_{j+k}, \quad
\sigma_j {\overline {\eta}}_i {\sigma}^{-1}_j= {\eta}_{j-i-1}-
{\eta}_{j-i}.
\label{auto1can}
\end{eqnarray}
Let us mention that besides the coordinates ${\eta}_j$ and
${\overline {\eta}}_{j}$, there is one more set of the canonical
coordinates ${\widetilde {\eta}}_j$ and ${\widetilde {\overline
{\eta}}}_{j}$ with the lattice--local equations of motion, related with the
former ones by the transformation
\begin{eqnarray}
{\widetilde {\eta}}_j={\eta}_{j-1}-{\eta}_{j}, \quad {\widetilde {\overline
{\eta}}}_{j}=\sum_{k=1}^{\infty} {\overline {\eta}}_{j-k} \quad
\Leftrightarrow \quad {\overline {\eta}}_j=-{\widetilde {\overline
{\eta}}}_{j}+ {\widetilde {\overline {\eta}}}_{j+1}, \quad
{\eta}_j=-\sum_{k=0}^{\infty} {\widetilde {\eta}}_{j-k},
\label{transcan2}
\end{eqnarray}
which does not include the fields ${\xi}_j$ and ${\overline {\xi}}_{j}$.
In addition to this, of course there exist other canonical
transformations of the Poisson algebra \p{hamstrcan1} which mix all fields
and also lead to local equations for them.

Following the standard procedure, one can derive the Lagrangian ${\cal L}$
and action ${\cal S}$,
\begin{eqnarray}
&&{\cal S}=\int dt {\cal L}\equiv \int dt [\sum_{j=-\infty}^{\infty}(p_j
{\textstyle{\partial\over\partial t_1}}x_j  +
{\xi}_j{\textstyle{\partial\over\partial t_1}}{\overline {\xi}}_j
+ {\eta}_j{\textstyle{\partial\over\partial t_1}}{\overline {\eta}}_j) -
H_2] \label{lagr}\\
&&=\int dt\sum_{j=-\infty}^{\infty}[\frac{1}{2}
({\textstyle{\partial\over\partial t_1}}x_j)^2+
{\xi}_j{\textstyle{\partial\over\partial t_1}}{\overline {\xi}}_j
+ {\eta}_j{\textstyle{\partial\over\partial t_1}}{\overline {\eta}}_j-
e^{x_j-x_{j-1}}-e^{x_j} \xi_j{\overline \eta}_{j}-
e^{-x_{j-1}}({\eta}_{j-1} -{\eta}_{j}){\overline {\xi}}_{j}], ~ ~ ~ ~ ~
\nonumber
\end{eqnarray}
which can be important in connection with the quantization problem.
As usual, the variation of the action ${\cal S}$ with respect to the
fields $\{ x_{j}, {\xi}_j, {\overline {\xi}}_{j}, {\eta}_j,
{\overline {\eta}}_{j} \}$ produces the equations of motion \p{flow1can} 
for them, where the momenta $p_j$ are replaced by 
${\textstyle{\partial\over\partial
t_1}}x_j$. If, in addition to the momenta, the fields ${\eta}_j$ and
${\overline {\eta}}_j$ are also eliminated from eqs. \p{flow1can} by means
of corresponding equations expressing them in terms of the fields
$\{ x_{j}, {\xi}_j, {\overline {\xi}}_{j}\}$ and their $t_1$-derivatives,
the remaining equations become 
\begin{eqnarray}
&&{\textstyle{{\partial}^2\over{\partial} {t_1}^2}}x_j=e^{x_{j+1}-x_{j}}-
e^{x_j-x_{j-1}}-{\xi}_j
{\textstyle{\partial\over\partial t_1}}{\overline {\xi}}_j+
{\overline {\xi}}_{j+1}{\textstyle{\partial\over\partial t_1}}{\xi}_{j+1},
\nonumber\\&& {\textstyle{\partial\over\partial t_1}}
(e^{x_{j-1}}{\textstyle{\partial\over\partial t_1}}{\xi}_j)
=e^{x_{j-1}}{\xi}_{j-1} -e^{x_j}{\xi}_j, \quad
{\textstyle{\partial\over\partial t_1}}
(e^{-x_j}{\textstyle{\partial\over\partial t_1}}{\overline {\xi}}_j)
=e^{-x_{j}}{\overline {\xi}}_{j+1}-e^{-x_{j-1}}{\overline {\xi}}_{j}
\label{flow1restr}
\end{eqnarray}
and reproduce the set of the restricted f--Toda chain equations \cite{ls},
which form a subset of the f-Toda chain equations (see section 6).

The canonical basis \p{transcan1} is not important only for
the Lagrangian formulation, there is also another reason to introduce it. 
This is the known fact that the basis \p{transcan1} can be more
convenient than the old one \p{def} to study the reductions
\p{subsets}--\p{bc1} of some structures
characterizing the infinite $N=2$ supersymmetric Toda lattice hierarchy.
Indeed, for the case of the bosonic Toda lattice hierarchy, in \cite{mt}
it was argued that for the recursion operator such 
reduction is not admitted at all in the old basis $\{a_i, b_i\}$, whereas the
reduction is possible in the canonical basis
$\{x_i,p_i\}$ (in both bases the Hamiltonian structures can be reduced). 
One expects that similar arguments should be relevant also for the 
supersymmetric case.
Having this in mind, we present the second Poisson
structure of the N=2 super Toda lattice hierarchy in the canonical basis, 
\begin{eqnarray}
&& \{ x_i, x_j \}_2={\varepsilon}_{i,j}, \nonumber\\
&& \{ x_i, p_j \}_2=-p_i \delta_{i,j}, \nonumber\\
&& \{ x_i,{\xi}_j \}_2 =- \frac{1}{2} {\xi}_j 
({\varepsilon}_{i,j}+\delta_{i,j}), \nonumber\\
&& \{ x_i,{\overline {\xi}}_j \}_2=\frac{1}{2}{\overline {\xi}}_j
({\varepsilon}_{i,j}+\delta_{i,j}), \nonumber\\
&& \{ x_i,{\eta}_j \}_2 = \frac{1}{2}
({\eta}_j-{\eta}_i){\varepsilon}_{i,j}, \nonumber\\
&& \{ x_i,{\overline {\eta}}_j \}_2=-\frac{1}{2}{\overline {\eta}}_j
({\varepsilon}_{i,j}+\delta_{i,j}), \nonumber\\
&& \{ p_i, p_j \}_2=e^{x_i-x_{j}}\delta_{i,j+1}-e^{x_j-x_{i}}
\delta_{i,j-1}, \nonumber\\
&& \{ p_i,{\xi}_j \}_2 = -e^{-x_i}({\eta}_i-{\eta}_j)
\delta_{i,j-1},\nonumber\\
&& \{ p_i,{\overline {\xi}}_j \}_2=e^{x_i} {\overline {\eta}}_j
\delta_{i,j}, \nonumber\\
&& \{ p_i,{\eta}_j \}_2 = \frac{1}{2}e^{x_i} {\xi}_i 
({\varepsilon}_{i,j}-\delta_{i,j}), \nonumber\\
&& \{ p_i,{\overline {\eta}}_j \}_2 =-e^{-x_i} 
{\overline {\xi}}_j\delta_{i,j-1}, \nonumber\\
&& \{ {\xi}_i, {\eta}_j \}_2 =
-\frac{1}{2}{\xi}_i({\eta}_j-{\eta}_i)
{\varepsilon}_{i,j}, \nonumber\\
&& \{ {\xi}_i, {\overline {\eta}}_j \}_2=\frac{1}{2}
{\xi}_i {\overline {\eta}}_j ({\varepsilon}_{i,j}-\delta_{i,j})
-e^{-x_j}\delta_{i,j+1}, \nonumber\\
&&\{{\overline {\xi}}_i, {\eta}_j\}_2=
-\frac{1}{2}({\eta}_{j}-{\eta}_i)
{\overline {\xi}}_i({\varepsilon}_{i,j}-\delta_{i,j})-
\frac{1}{2}e^{x_i}({\varepsilon}_{i,j}+\delta_{i,j}), \nonumber\\
&& \{ {\overline \xi}_i , {\overline {\eta}}_j \}_2 =
-\frac{1}{2}{\overline \xi}_i{\overline {\eta}}_j
({\varepsilon}_{i,j}-\delta_{i,j}),  \nonumber\\
&& \{ {\eta}_i, {\overline {\eta}}_j \}_2=\frac{1}{2}
{\xi}_j{\overline {\xi}}_j({\varepsilon}_{i,j}+\delta_{i,j})
+\frac{1}{2}p_j({\varepsilon}_{i,j}-\delta_{i,j}),
\label{canhamstr2}
\end{eqnarray}
where we have introduced the antisymmetric lattice
function ${\varepsilon}_{i,j}$,
\begin{eqnarray}
{\varepsilon}_{i,j} = -{\varepsilon}_{j,i} \equiv 1, \quad 
if \quad   i > j  
\label{eps}
\end{eqnarray}
with the evident properties:
\begin{eqnarray}
{\varepsilon}_{-i,-j}=-{\varepsilon}_{i,j}, \quad 
{\varepsilon}_{i,j-1}-{\varepsilon}_{i,j}=
{\delta}_{i,j-1}+{\delta}_{i,j},   
\label{eps1}
\end{eqnarray}
and only nonzero brackets are written down. Together with the inverse
matrix for the first Hamiltonian structure \p{hamstrcan1}, which can be
easily derived due to the very simple structure of those equations, the 
relations \p{canhamstr2} define the recursion operator in the canonical
basis \p{transcan1}. Its explicit expression can be easily obtained and will
not be written down here.

\end{document}